\documentclass[10pt,journal]{IEEEtran}
\usepackage{amssymb}
\usepackage{amsmath}
\usepackage{cite}
\usepackage{url}
\usepackage{xcolor}
\usepackage{cite,graphicx,amsmath,amssymb}
\usepackage{subfigure}
\usepackage{fancyhdr}
\usepackage{mdwmath}
\usepackage{mdwtab}
\usepackage{caption}
\usepackage{amsthm}
\usepackage{setspace}
\usepackage{hyperref}
\usepackage{algorithm}
\usepackage{algorithmic}
\usepackage{pdfpages}
\usepackage{mathtools}

\usepackage{bm}
\hypersetup{colorlinks=false}

\graphicspath{{./src}}
\usepackage[acronym,nogroupskip,nonumberlist,nopostdot]{glossaries}
\makeglossaries
\newacronym{aod}{AOD}{angle of departure}
\newacronym{aoa}{AOA}{angle of arrival}
\newacronym{snr}{SNR}{signal-to-noise ratio}
\newacronym{6g}{6G}{sixth-generation}
\newacronym{5g}{5G}{fifth-generation}
\newacronym{sre}{SRE}{smart radio environment}
\newacronym{stars}{STAR-RISs}{simultaneously transmitting and reflecting reconfigurable intelligent surfaces}
\newacronym{star}{STAR-RIS}{simultaneously transmitting and reflecting reconfigurable intelligent surface}
\newacronym{mimo}{MIMO}{multiple-input multiple-multiple}
\newacronym{cscg}{CSCG}{circularly symmetric complex Gaussian}
\newacronym{los}{LoS}{line-of-sight}
\newacronym{em}{EM}{electromagnetic}

\newtheorem{remark}{Remark}
\newtheorem{theorem}{Theorem}

\newtheorem{lemma}{Lemma}

\newtheorem{corollary}{Corollary}

\allowdisplaybreaks

\title{Near-Field Integrated Sensing, Positioning, and Communication: A Downlink and Uplink Framework}

\author{

Haochen~Li,~\IEEEmembership{Graduate Student Member,~IEEE,}
Zhaolin~Wang,~\IEEEmembership{Graduate Student Member,~IEEE,}
Xidong~Mu,~\IEEEmembership{Member,~IEEE,}
Pan~Zhiwen,~\IEEEmembership{Member,~IEEE,}
Yuanwei~Liu,~\IEEEmembership{Senior Member,~IEEE,}

\thanks{Part of this work has been submitted to the IEEE International Conference on Communications, in 2024~\cite{Li}.}
\thanks{Haochen~Li and Pan~Zhiwen are with National Mobile Communications Research Laboratory, Southeast University, Nanjing 210096, China, and also with Purple Mountain Laboratories, Nanjing 211100, China (email: lihaochen@seu.edu.cn, pzw@seu.edu.cn).}
\thanks{Zhaolin Wang, Xidong Mu, and Yuanwei Liu are with the School of Electronic Engineering and Computer Science, Queen Mary University of London, London E1 4NS, U.K. (e-mail: zhaolin.wang@qmul.ac.uk, xidong.mu@qmul.ac.uk, yuanwei.liu@qmul.ac.uk).}
}
\begin{document}

\maketitle

\begin{abstract}
    A near-field integrated sensing, positioning, and communication (ISPAC) framework is proposed, where a base station (BS) simultaneously serves multiple communication users and carries out target sensing and positioning. A novel double-array structure is proposed to enable the near-field ISPAC at the BS. Specifically, a small-scale assisting transceiver (AT) is attached to the large-scale main transceiver (MT) to empower the communication system with the ability of sensing and positioning. Based on the proposed framework, the joint angle and distance Cram\'er-Rao bound (CRB) is first derived. Then, the CRB is minimized subject to the minimum communication rate requirement in both downlink and uplink ISPAC scenarios: 1) For downlink ISPAC, a downlink target positioning algorithm is proposed and a penalty dual decomposition (PDD)-based double-loop algorithm is developed to tackle the non-convex optimization problem. 2) For uplink ISPAC, an uplink target positioning algorithm is proposed and an efficient alternating optimization algorithm is conceived to solve the non-convex CRB minimization problem with coupled user communication and target probing design. Both proposed optimization algorithms can converge to a stationary point of the CRB minimization problem. Numerical results show that: 1) The proposed ISPAC system can locate the target in both angle and distance domains merely relying on single BS and limited bandwidths; and 2) the positioning performance achieved by the hybrid-analog-and-digital ISPAC approaches that achieved by fully digital ISPAC when the communication rate requirement is not stringent.
\end{abstract}

\begin{IEEEkeywords}
{C}ram\'er-Rao bound, sensing and positioning, near-field communications.
\end{IEEEkeywords}
\section{Introduction} \label{sec:introduction}
With ever-increasing performance targets for the next-generation communication, the upcoming sixth-generation (6G) communication is predicted to possess large-scale antenna arrays and work at significantly high frequencies~\cite{dang2020should}. This emerging trend brings changes to the electromagnetic (EM) properties of the wireless environment. Specifically, the EM region around base stations (BSs) can be divided into near-field and far-field regions, demarcated by the Rayleigh distance~\cite{kraus2002antennas}. In the far-field region, EM waves can be approximated as planar waves, while in the near-field region, EM waves necessitate precise modeling as spherical waves~\cite{10220205}. The Rayleigh distance increases with the array aperture and communication frequency~\cite{1137900}. With large-scale antenna arrays and high frequencies, the Rayleigh distance in 6G communication systems can span several tens to hundreds of meters, which leads to communications taking place in the near-field region. 

The transition from traditional far-field communication (FFC) to near-field communication (NFC) presents new opportunities for communication design. On the one hand, 6G communication demands further enhancement of capacity to support new services such as ultra-high-definition video streaming and extended reality (XR)~\cite{zhang20196g}. Near-field channels encompass both angular and distance domains, which provide enhanced degrees of freedom (DoFs) for communication design, thereby elevating multiplexing gain and connectivity of communication systems. On the other hand, driven by the ambitious ``ubiquitous wireless intelligence’’ goal of 6G communication, communication systems need to be aware of surrounding environments to support new applications such as smart city and Metaverse~\cite{rajatheva2020white}. The distance and angle information embedded in the near-field channel can be utilized to support sensing targets in the near-field region and obtaining target positions, i.e., sensing and positioning, without necessitating larger bandwidth resources or cooperation between multiple BSs~\cite{5571900}.\vspace{-0.1cm}
\subsection{Prior Works}
\subsubsection{Near-field communication}
The potential benefits of NFC have been investigated in many prior works~\cite{zhang2022beam,9620081,10123941}. Exploiting the angle and distance domain contained in the near-field channel, NFC can achieve spotlight-like beamfocusing and thus concentrate communication signals on the locations of intended users. Near-field multiple input multiple output (MIMO) communications were investigated in~\cite{zhang2022beam}, which showed that the near-field beamfocusing can enhance the spatial multiplexing gain in multi-user systems. Based on this observation, the spectral efficiency of near-field MIMO communications under three basic precoding strategies was analyzed in~\cite{9620081}. It was demonstrated that the new distance domain emerging in the near-field region provides extra DoFs for inter-user interference (IUI) management. Moreover, the authors of~\cite{10123941} proposed the new concept of location division multiple access and proved the asymptotic orthogonality of near-field beamfocusing vectors.
\subsubsection{Target sensing and positioning}
To achieve target sensing and positioning, the BS needs to extract position-related information, such as received signal strength (RSS), angle-of-arrival (AoA), time-of-arrival (ToA), or time-difference-of-arrival (TDoA)~\cite{xiao2022overview}. However, the accuracy of RSS is limited due to the absence of a precise model for the relationship between RSS and propagation distance~\cite{wang2011new}. Besides, cooperation between multiple BSs is required to infer the target position using the AoAs~\cite{wang2015asymptotically}. Furthermore, using time-based metrics, i.e., ToA and TDoA, for target positioning requires synchronization between multiple BSs, which introduces additional challenges in the design of positioning systems. Accurate and reliable time-based metrics require large bandwidth resources since fine delay resolution is needed~\cite{5571900}. In the near-field, both the angle and distance information are available in the narrow band channels thanks to the spherical-wave propagation. Thus, target positioning through single BS and limited bandwidths can be realized. The near-field target positioning with large-scale MIMO is investigated in work~\cite{qiao2023sensing}, where the large array at the BS is divided into several sub-arrays to estimate user locations via multi-subarray collaboration. The Cram\'er-Rao bound (CRB) is used as the performance metric for target positioning as it provides the lower bound for the estimations of the target angle and distance. In~\cite{wang2023cram}, the angle and distance CRBs are derived for the near-field MIMO radar system for both monostatic and bistatic models. More generally, the joint location CRB for near-field positioning concerning multiple targets was given in~\cite{hua2023near}.\vspace{-0.1cm}
\subsection{Motivations and Contributions}
Based on the aforementioned works, it can be observed that the application of near-field characteristics is beneficial to both user communication and target sensing. We naturally want to propose an integrated sensing, positioning, and communication (ISPAC) framework. However, there exist the following challenges:
\begin{itemize}
\item If following the conventional array structure used in far-field ISAC, the BS in the ISPAC framework requires a large-scale full-duplex antenna for simultaneous signal transmission and reception~\cite{wang2023near}. This necessity leads to high power consumption and construction costs~\cite{cong2023near}, which motivates us to propose a new array structure to reduce the complexity of the BS while retaining the positioning function.
\item Near-field target positioning requires locating the target in both angle and distance domains. When using traditional algorithms like the MUltiple SIgnal Classification (MUSIC) algorithm or maximum likelihood estimation (MLE) algorithm to carry out target positioning the time-consuming two-dimensional (2D) search is required. This motivates us to develop new positioning algorithms that are more efficient and suitable for near-field ISPAC systems.
\end{itemize}

Driven by the above challenges, we propose a novel near-field ISPAC framework, where the target sensing and positioning is simultaneously carried out with communication. Our main contributions are summarized as follows:
\begin{itemize}
\item We propose a novel double-array ISPAC BS structure, where an assisting transceiver (AT) is installed to the main transceiver (MT) for seamlessly integrating the sensing function into the pre-existing NFC networks without necessitating complex modifications to the hardware. Based on this setup, the near-field joint angle and distance CRB is derived for target positioning. The CRB minimization problems for both downlink and uplink cases are formulated under the commutation quality of service (QoS) constraint and the HAD precoding constraint.
\item For downlink ISPAC, a two-stage downlink positioning algorithm is proposed to successively estimate the target angle and distance with low-complex one-dimensional (1D) search and circumvent the complex 2D search required by traditional positioning algorithms. Specifically, the MUSIC algorithm is adopted to discriminate the target angle. With the estimated angle, the MLE is used for target distance estimation. The complicated CRB is reformulated into an equivalent form with the Fisher information matrix (FIM). A penalty dual decomposition (PDD)-based algorithm is conceived to tackle the non-convex optimization problem.
\item For uplink ISPAC, a two-stage uplink positioning algorithm is proposed, where the target angle and distance parameters in the sensing channel are split and then successively estimated with the MUSIC algorithm. An alternating optimization (AO) algorithm is conceived for CRB minimization, where the HAD receiving precoders at MT, and the probing covariance matrix at the AT are alternately obtained.
\item Our numerical results confirm the convergence and effectiveness of the proposed algorithms. It is proved that the proposed positioning algorithms can accurately estimate the target angle and distance. It is also verified that the proposed ISPAC framework can guarantee the QoS of communication users while supporting target positioning. Besides, we demonstrate that more DoFs are available for sensing and positioning design when near-field beamfocusing is used for communication design rather than conventional far-field beamsteering.
\end{itemize}

The remainder of this paper is organized as follows: In Section II, the system setup, channel models, and signal models for the ISPAC framework are introduced.  In Section III, we present a target location estimation algorithm customized for the downlink ISPAC and provide a solution to the CRB minimization problem. Section IV presents the target location estimation algorithm designed for the uplink ISPAC scenario and addresses the associated CRB minimization problem. In Section V, simulation results are presented to validate the efficacy of the proposed algorithms. Section VI concludes this paper.

\indent \textit{Notations:} Lowercase letters, lowercase bold letters, and capital bold letters denote scalars, vectors, and matrices, respectively. The $M \times K$ dimensional complex matrix space is denoted by $\mathbb{C}^{M \times K}$. The superscripts $(\cdot)^\mathrm{T}$, $(\cdot)^\mathrm{*}$, and  $(\cdot)^\mathrm{H}$ represent the operations of transpose, conjugate, and conjugate transpose, respectively. $\text{diag}\left( \cdot \right)$ and $\text{Bdiag}\left( \cdot \right)$ denote the diagonal and block diagonal operations, respectively. $\text{tr}(\mathbf{A})$ and $\text{rank}(\mathbf{A})$ denote the trace and the rank
of matrix $\mathbf{A}$, respectively. The distribution of a circularly symmetric complex Gaussian (CSCG) random vector with zero mean and covariance matrix $\mathbf{A}$ is denoted as $\mathcal{C N}\left(\mathbf{0}, \mathbf{A}\right)$. \vspace{-0.2cm}

\section{System Model} \label{sec:system_model}
\begin{figure}[ht]
\centering
\subfigure[Illustration of the downlink ISPAC framework.]{\label{fig:downlink}
\includegraphics[width=3.1in]{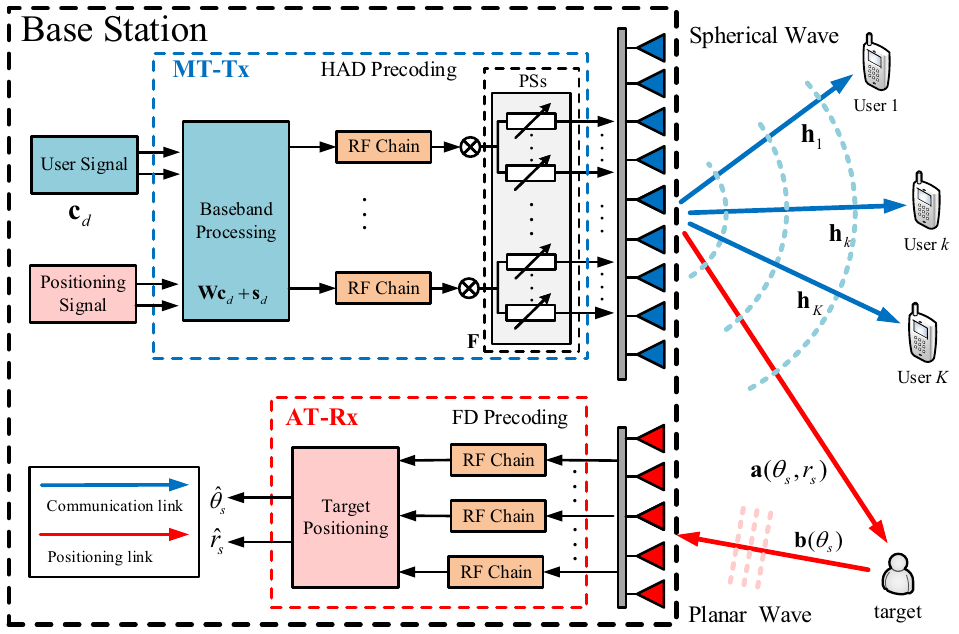}}\vspace{-0.2cm}
\subfigure[Illustration of the uplink ISPAC framework.]{\label{fig:uplink}
\includegraphics[width=3.1in]{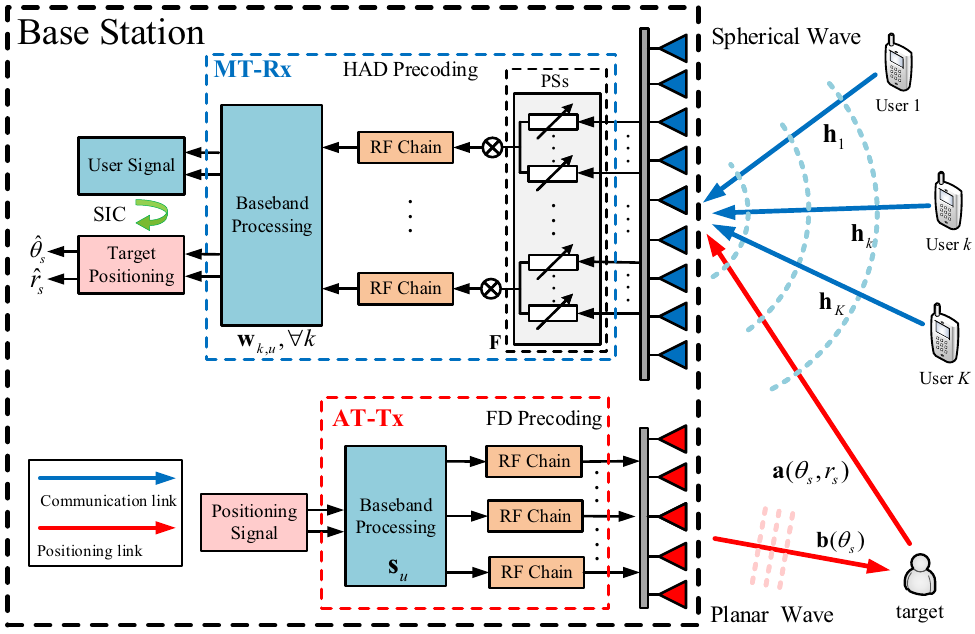}}\vspace{-0.2cm}
\caption{The proposed near-field integrated sensing, positioning, and communication framework.}\label{fig:system_model}\vspace{-0.6cm}
\end{figure}

As shown in Fig.~\ref{fig:system_model}, we propose a near-field ISPAC framework, where a BS equipped with a $N_a$-antenna MT and a $N_b$-antenna AT serves $K$ single-antenna communication users and locates a target. The MT adopts a large-scale array to cover communication users and the target in its near-field region. To reduce the power consumption and hardware complexity of the MT, the partially-connected HAD precoding architecture is adopted. The MT is equipped with $N_{RF}$ radio frequency~(RF) chains and every RF chain is attached to a $M$-antenna sub-array with $M=N_a/N_{RF}$ (We assume $N_a/N_{RF}$ is an integer for brevity). With this HAD precoding architecture, the analog precoder of the MT at the BS is given by 
\begin{equation}\label{eqn:analog}
    \mathbf{F} = \frac{1}{\sqrt{M}}\text{Bdiag}\left({\bm{f}}_1,{\bm{f}}_2,\cdots,{\bm{f}}_{{N_{RF}}}\right),
\end{equation}
where ${\bm{f}}_i\in \mathbb{C}^{M \times 1}$, for $i=1,2,\cdots,N_{RF}$. $\mathbf{F}$ belongs to a block matrices set $\mathcal{A}_F$. The diagonal of each block in $\mathbf{F}$, i.e., ${\bm{f}}_i$,  is a $M$ dimension vector whose elements have the same amplitude~$1$. On the contrary, the AT adopts fully digital (FD) structure as its antenna number is moderate.\vspace{-0.2cm}
\subsection{Channel Model}
In the following, the near-field communication channel vectors $\mathbf{h}_k, \forall k \in \mathcal{K}=\{1,2,\cdots,K\}$ and the mixed far and near-field sensing channel matrix $\mathbf{G}$ are introduced.
\subsubsection{The near-field and far-field array response vectors}
Suppose the reference point of a $N$-antenna ULA is located at $\left(0, 0\right)$ in the $XY$-plane, and then the coordinate of the $n$-th antenna is $\left(0, \left(n-1\right)d\right)$. A single-antenna user is located at $\left(x, y\right)$, whose polar coordinate is $\left(\theta, r\right)=\left(\text{atan} (\frac{y}{x}),\sqrt{x^2+y^2}\right)$. The line-of-sight (LoS) channel between the $n$-th antenna of the ULA and the single-antenna user can be represented as~\cite{10273424}\vspace{-0.1cm}
\begin{equation}
\begin{aligned}
\left[\mathbf{h}\right]_n=\alpha_n e^{-jk_cr_n},
\end{aligned} \vspace{-0.1cm}
\end{equation}
where $k_c=2\pi /\lambda_c$ is the wavenumber and $\lambda_c$ is the wavelength. $\alpha_n$ and $r_n$ represent the path loss and distance between the $n$-th antenna of the ULA and the single-antenna user. Based on the Fresnel approximation~\cite{7942128}, we can assume the path loss for the channel between all  antennas and the user is the same, i.e., $\alpha_n=\alpha, \forall n$. Then we have\vspace{-0.1cm}
\begin{equation}\label{eqn:array_response_vector}
\mathbf{h}=\alpha\left[e^{-jk_cr_1}, e^{-jk_cr_2}, \cdots, e^{-jk_cr_N}\right]^\mathrm{T}=\alpha\mathbf{e}^N\left(\theta,r\right).\vspace{-0.1cm}
\end{equation}
where $\mathbf{e}^N\left(\theta,r\right)$ is the array response vector of the $N$-antenna ULA. Using the first-order Taylor expansion, $r_n$ can be approximated as $r_n\approx r-\left(n-1\right)d\sin\theta$. Then, we have the following approximation\vspace{-0.1cm}
\begin{equation}\label{eqn:farchannel}
\begin{aligned}
\mathbf{h}\approx& \mathbf{h}_\text{far}\left(\theta\right)=\alpha e^{-jk_cr}[1,e^{jk_c d\sin\theta},\cdots,\\[-0.05cm]
& e^{jk_c\left(N-1\right)d\sin\theta}]^\mathrm{T}=\alpha e^{-jk_cr}\mathbf{e}^N_\text{far}\left(\theta\right),
\end{aligned}\vspace{-0.1cm}
\end{equation}
where $\mathbf{e}^N_\text{far}\left(\theta\right)$ is the far-field array response vector of the $N$-antenna ULA~\cite{6717211}. However, the channel model~\eqref{eqn:farchannel} is insufficient to capture the characteristic of spherical wave in near field. By using the second-order Taylor expansion, $r_n$ can be approximated as $r_n\approx r-\left(n-1\right)d\sin\theta+\delta_n$, where $\delta_n={\left(n-1\right)^2d^2\cos^2\theta}/{2r}$. Then we have following approximation\vspace{-0.1cm}
\begin{equation}
\begin{aligned}\label{eqn:nearchannel}
\mathbf{h}\approx&\mathbf{h}_\text{near}\left(\theta,r\right)
=\alpha e^{-jk_cr}[1,e^{jk_c (d\sin\theta-\delta_2)},\cdots,\\[-0.05cm]
& e^{jk_c\left((N-1)d\sin\theta-\delta_n\right)}]^\mathrm{T}
=\alpha e^{-jk_cr}\mathbf{e}^N_\text{near}\left(\theta,r\right),
\end{aligned}\vspace{-0.1cm}
\end{equation}
where $\mathbf{e}^N_\text{near}\left(\theta,r\right)$ is the near-field array response vector of the $N$-antenna ULA~\cite{zhang2022beam}.

\subsubsection{The near-field communication channel model}
Due to the large aperture of the MT, the channels between the MT and the communication users, i.e., $\mathbf{h}_k, \forall k \in \mathcal{K}$, should be modeled with the near-field channel model. 

Taking the point scatterer assumption~\cite{9598863}, the channel from user $k$ to the MT consists of both the LoS part $\mathbf{h}^{k}_0$ and the non-line-of-sight (NLoS) part $\sum_{l=1}^{L_k}\mathbf{h}^{k}_l$ introduced by $L_k$ scatterers, i.e.,\vspace{-0.1cm}
\begin{equation}\label{channel_h_k}
\!\!\mathbf{h}_k\!\!=\!\!\sum\nolimits_{l=0}^{L_k}\!\!\mathbf{h}^{k}_l\!\!=\!\alpha_k\mathbf{e}^{N_a}_\text{near}\!\!\left(\theta_k,r_k\right)\!+\!\!\sum\nolimits_{l=1}^{L_k}\!\!\frac{\alpha_l^k}{\sqrt{L_k}}\mathbf{e}^{N_a}_\text{near}\!\!\left(\theta_l^k,r_l^k\right),\vspace{-0.1cm}
\end{equation} 
where $\mathbf{h}^{k}_0$ denotes the LoS part of the channel, while $\mathbf{h}^{k}_l$ stands for the NLoS part of the channel corresponding to the $l$-th scatterer. $\theta_k(\theta_l^k)$ and $r_k(r_l^k)$ represent the angle and distance of the user ($l$-th scatterer) with respect to the reference point of the MT array. With $\tilde{r}_l^k$ representing the distance between user $k$ and the $l$-th scatterer associated to it, $\alpha_k=\tilde{\alpha}_ke^{-jk_cr_k}$ and $\alpha_l^k=\tilde{\alpha}_l^ke^{-jk_c(r_l^k+\tilde{r}_l^k)}$ are the equivalent path loss of the LoS channel and the $l$-th NLoS channel, respectively. Specifically, $\tilde{\alpha}_k$ mainly accounts for the free-space path loss, while $\tilde{\alpha}_l^k$ is determined by both the free-space path loss and the reflection coefficient of the $l$-th scatter associated with user~$k$.

\subsubsection{The mix-field sensing channel model}
For positioning the target, BS sends probing signal to the target and then gathers the echo signals. Using the array response vector in \eqref{eqn:array_response_vector}, the two hop AT$\rightarrow$target$\rightarrow$MT channel can be expressed as:\vspace{-0.1cm}
\begin{equation}\label{channel_probe}
\mathbf{G} = {\beta}_r {\beta}_p^2 \mathbf{e}^{N_a}(\theta_s, r_s) (\mathbf{e}^{N_b}(\theta_s, r_s))^T,\vspace{-0.1cm}
\end{equation}
where $(\theta_s, r_s)$ is the polar coordinate of the target. ${\beta}_r$ and ${\beta}_p$ represent the reflection coefficient and the path loss between the BS and target, respectively. 

The MT has a large aperture due to the large antenna number, while the AT only contains a limited number of antennas. We assume the target is located in the near-field of the MT and the far-field of the AT. Hence, the approximation in \eqref{eqn:nearchannel} and \eqref{eqn:farchannel} can be used to approximated terms $\mathbf{e}^{N_a}(\theta_s, r_s)$ and $\mathbf{e}^{N_b}(\theta_s, r_s)$ in \eqref{channel_probe}, respectively. Then, the mix-field target sensing channel can be expressed as:\vspace{-0.1cm}
\begin{equation}\label{eqn:G}
\mathbf{G}  = \beta_s \mathbf{e}_\text{near}^{N_a}(\theta_s, r_s) \mathbf{e}_\text{far}^{N_b}(\theta_s)= \beta_s \mathbf{a}(\theta_s, r_s) \mathbf{b}^T(\theta_s),\vspace{-0.1cm}
\end{equation}
where $\beta_s$ is the equivalent path loss accounting for the reflection factor and the two hop path loss.
\subsection{Signal Model}
\subsubsection{Downlink ISPAC}
When the BS serves downlink communication users, the \emph{downlink ISPAC working mode} is adopted to enable the target positioning, where the MT and AT work as transmitter and receiver, respectively. As shown in Fig.~\ref{fig:downlink}, the MT sends communication and probing signals simultaneously, covering all the communication users and target in its near-field region. The AT collects the echo signal reflected by the target. At time slot $t$, the signal sent by the BS can be expressed as\vspace{-0.1cm}
\begin{equation}
    \mathbf{x}(t) = \mathbf{F} \sum\nolimits_{k \in \mathcal{K}} \mathbf{w}_{k,d} {c}_{k,d}(t)+\mathbf{F}\mathbf{s}_d(t),\vspace{-0.1cm}
\end{equation}
where $\mathbf{F}$ is the MT analog precoder as defined in $\eqref{eqn:analog}$. $c_{k,d}(t)$ is the downlink communication signal for user $k$ with normalized power. $\mathbf{w}_{k,d}\in \mathbb{C}^{N_{RF}\times 1}$ is the digital precoder for user $k$. $\mathbf{s}_d(t) \in \mathbb{C}^{N_{RF}\times 1}$ is the probing signal with covariance matrix $\mathbf{R}_{d}=\mathbb{E} [{\mathbf{s}_d(t)} {\mathbf{s}_d^H(t)} ]$. The covariance matrix of $\mathbf{x}(t)$ can be calculated as \vspace{-0.1cm}
\begin{equation}\label{RX}
    \mathbf{R}_x = \mathbf{F}\mathbf{W}\mathbf{W}^H\mathbf{F}^H+\mathbf{F}\mathbf{R}_{d}\mathbf{F}^H = \mathbf{F}\tilde{\mathbf{R}}_x\mathbf{F}^H,\vspace{-0.1cm}
\end{equation}
where $\mathbf{W}=\left[ {{\bf{w}}_{1,d},{\bf{w}}_{2,d}, \cdots ,{\bf{w}}_{K,d}} \right]$ is the downlink precoding matrix and $\tilde{\mathbf{R}}_x=\mathbf{W}\mathbf{W}^H+\mathbf{R}_{d}$. In practice, $\mathbf{R}_x$ can be approximated by the average covariance matrix over $T$ time slot, i.e., $\mathbf{R}_x \approx \frac{1}{T}\mathbf{X}\mathbf{X}^H$, where $\mathbf{X}=\left[ {{\bf{x}}(1),{\bf{x}}(2), \cdots ,{\bf{x}}(T)} \right]$. With the transmitted signal ${\bf{x}}(t)$, the received signal at user $k$ at time slot $t$ is\vspace{-0.1cm}
\begin{equation}
    {y}_k(t) = \mathbf{h}_k^T\mathbf{F}\mathbf{W}\mathbf{c}_{d}(t)+ \mathbf{h}_k^T\mathbf{F}\mathbf{s}_d(t)+{n}_k(t),\vspace{-0.1cm}
\end{equation}
where $\mathbf{c}_d(t)=\left[ c_{1,d}(t),c_{2,d}(t), \cdots ,c_{K,d}(t)\right]$ is the downlink signal vector and ${n}_k(t) \sim \mathcal{CN}({0}, \sigma_0^2)$ is the additive white Gaussian noise at user $k$. The SINR of user $k$ is given by \vspace{-0.1cm}
\begin{equation}
    \text{SINR}_{k,d} = \frac{|\mathbf{h}_{k}^T\mathbf{F}\mathbf{w}_{k,d}|^2}{\sum_{i\neq k}|\mathbf{h}_{k}^T\mathbf{F}\mathbf{w}_{i,d}|^2+\mathbf{h}_{k}^T\mathbf{F}\mathbf{R}_{d}\mathbf{F}^H\mathbf{h}_{k}^*+\sigma_0^2}.\vspace{-0.1cm}
\end{equation}

The echo signal received by the BS at time slot $t$ can be given by\vspace{-0.1cm}
\begin{equation}
    \mathbf{y}_{s,d}(t) = \mathbf{G}^T\mathbf{x}(t)+\mathbf{H}_{SI}^T\mathbf{x}(t) +\mathbf{n}_d(t),\vspace{-0.1cm}
\end{equation}
where $\mathbf{n}_d(t) \sim \mathcal{CN}(\mathbf{0}, \sigma_d^2 \mathbf{I}_{N_b})$ denotes the complex Gaussian noise at the receiver. $\mathbf{H}_{SI}\in \mathbb{C}^{N_a\times N_b}$ is the self-interference (SI) channel from the AT to the MT. Assuming perfect SI cancellation, the echo signal collected at the BS over $T$ coherent time slot is $\mathbf{Y}_{s,d} = \mathbf{G}^T\mathbf{X}+\mathbf{N}_d$, where $\mathbf{Y}_{s,d}=\left[ {{\bf{y}}_{s,d}(1),{\bf{y}}_{s,d}(2), \cdots ,{\bf{y}}_{s,d}(T)} \right]$ and ${\mathbf{N}}_d=\left[ {\mathbf{n}}_d(1),{\mathbf{n}}_d(2), \cdots ,{\mathbf{n}}_d(T) \right]$. Then, the location of the target is estimated from $\mathbf{Y}_{s,d}$ with proposed two-stage downlink positioning algorithm, as elaborated in the next section.

\subsubsection{Uplink ISPAC} When the BS serves uplink communication users, the \emph{uplink ISPAC working mode} is adopted to enable simultaneous target positioning. Different from the downlink ISPAC working mode, the MT works as a receiver to collect uplink communication signals and echo probing signals. The AT sends probing signal for target positioning. The uplink communication signals are decoded under the interference of the echo signals, while the target positioning is carried out without interference from communication signals after successive interference cancelation (SIC).

As shown in Fig.~\ref{fig:uplink}, during the uplink ISPAC, the BS receives three superimposed signals, i.e. the communication signals for $K$ uplink users, the echo signal from the target, and the SI from the transmitter. The received signal at the BS at time slot $t$ is \vspace{-0.1cm}
\begin{equation}
    \!\mathbf{y}_u(t) \!= \!\mathbf{G} \mathbf{s}_u(t)\!+\sqrt{P_u}\sum\nolimits_{k}\! \mathbf{h}_kc_{k,u}(t)\!+\mathbf{H}_{SI}\mathbf{s}_u(t)\!+\mathbf{n}_u(t),\vspace{-0.1cm}
\end{equation}
where $P_u$ is the maximum transmitting power for each uplink user. $c_{k,u}(t) \in \mathbb{C}$ is the communication signal from user $k$ with normalized power. $\mathbf{n}_u(t) \sim \mathcal{CN}(\mathbf{0}, \sigma_u^2 \mathbf{I}_{N_a})$ denotes the complex Gaussian noise at the large-scale HAD receiver. $\mathbf{s}_u(t) \in \mathbb{C}^{N_b\times 1}$ is the probing signal transmitted by the small-scale FD transmitter. 

The covariance matrix of the probing signal is $\mathbf{R}_{u}=\mathbb{E} [{\mathbf{s}_u(t)} {\mathbf{s}_u^H(t)} ]$, which can be approximated by the average covariance matrix over $T$ time slot in practice, i.e., $\mathbf{R}_{u} \approx \frac{1}{T}\mathbf{S}_u\mathbf{S}_u^H$, where $\mathbf{S}_u=\left[ {{\bf{s}}_u(1),{\bf{s}}_u(2), \cdots ,{\bf{s}}_u(T)} \right]$. Assuming the perfect SI cancellation, the received signal after analog combining at the receiver at time slot $t$ is\vspace{-0.2cm}
\begin{equation}
    \tilde{\mathbf{y}}_u(t) =\mathbf{F}^H\mathbf{G} \mathbf{s}_u(t)+\sqrt{P_u}\mathbf{F}^H\sum\nolimits_{k} \mathbf{h}_kc_{k,u}(t)+\tilde{\mathbf{n}}_u(t),\vspace{-0.2cm}
\end{equation}
where $\tilde{\mathbf{n}}_u(t)= \mathbf{F}^H{\mathbf{n}_u}(t) \sim \mathcal{CN}(\mathbf{0}, \sigma_u^2 \mathbf{I}_{N_{RF}})$. Adopting linear digital combiner $\mathbf{w}_{k,u} \in \mathbb{C}^{N_{RF} \times 1}$ at the receiver, the effective signal for decoding $c_{k,u}(t)$ from $\tilde{\mathbf{y}}_u(t)$ is given by \vspace{-0.1cm}
\begin{equation} \label{eqn:signal}
\begin{aligned}
    \hat{c}_{k,u}(t)=&\sqrt{P_u} \mathbf{w}_{k,u}^H \mathbf{F}^H \mathbf{h}_{k} c_{k,u}(t)+ \mathbf{w}_{k,u}^H \mathbf{F}^H\mathbf{G} \mathbf{s}_u(t) \\
    &+ \mathbf{w}_{k,u}^H \mathbf{F}^H \sum\nolimits_{ k } \sqrt{P_u} \mathbf{h}_{i} c_{i}(t) + \mathbf{w}_{k,u}^H\tilde{\mathbf{n}}_u(t).\vspace{-0.1cm}
\end{aligned}
\end{equation} 
Then, the SINR of user $k$ is given by \vspace{-0.1cm}
\begin{equation} \label{eqn:rate_user_EN}
    \text{SINR}_{k,u} = \frac{P_u |\mathbf{w}_{k,u}^H \mathbf{F}^H \mathbf{h}_{k}|^2 }{\mathbf{w}_{k,u}^H \mathbf{F}^H \mathbf{R}_{k,u} \mathbf{F} \mathbf{w}_{k,u} },\vspace{-0.1cm}
\end{equation}
where the matrix $\mathbf{R}_{k,u}$ denotes the interference plus noise covariance matrix which can be calculated as follows:\vspace{-0.1cm}
\begin{equation}
    \mathbf{R}_{k,u} = \sum\nolimits_{i\neq k} P_u \mathbf{h}_{i} \mathbf{h}_{i}^H + \mathbf{G} \mathbf{R}_{u} \mathbf{G}^H + \sigma_u^2 \mathbf{I}_{N_a}.\vspace{-0.1cm}
\end{equation}
After decoding the communication signals at the receiver, the SIC technique is utilized to remove the effect of uplink communication data from the received signals. Thus, the effective signal for positioning the desired target at time slot $t$ is given by\vspace{-0.1cm}
\begin{equation}\label{eqn:ys}
    \mathbf{y}_{s,u}(t) = \mathbf{F}^H \mathbf{G} \mathbf{s}_u(t)+\tilde{\mathbf{n}}_u(t).\vspace{-0.1cm}
\end{equation}
Over $T$ coherent time slot, the received echo signals can be expressed as $\mathbf{Y}_{s,u} = \mathbf{F}^H \mathbf{G} \mathbf{S}_u+{\mathbf{N}_u}$, where $\mathbf{Y}_{s,u}=\left[ \mathbf{y}_{s,u}(1),\mathbf{y}_{s,u}(2), \cdots ,\mathbf{y}_{s,u}(T) \right]$ and ${\mathbf{N}_u}=\left[ \tilde{\mathbf{n}}_u(1),\tilde{\mathbf{n}}_u(2), \cdots ,\tilde{\mathbf{n}}_u(T) \right]$. Then, the location of the target i.e., $\theta_s$ and $r_s$, is estimated from $\mathbf{Y}_{s,u}$ with proposed two-stage uplink positioning algorithm, as elaborated in Section~IV.

\section{Target Positioning and CRB Optimization Design for Downlink ISPAC}
In this section, we propose the two-stage downlink positioning algorithm for estimating the target location during downlink ISPAC. Then, we design the downlink ISPAC system based on the CRB.
\subsection{Positioning Algorithm for Downlink ISPAC}
{In terms of downlink target positioning, the two-stage downlink positioning algorithm is proposed to estimate the target angle and distance. To elaborate, in the \emph{first stage}, the MUSIC algorithm is adopted for target angle estimation. Over $T$ time slot, the covariance matrix of the received echo signal can be approximated as $\mathbf{R}_{y,d}=\frac{1}{T}\mathbf{Y}_{s,d}\mathbf{Y}_{s,d}^H$. Using the eigenvalue decomposition, the noise space $\mathbf{U}_{n,d}$ with the dimension of $N_{b}-1$ is obtained. The received probing signal is spanned by the vector ${\mathbf{b}}(\theta)$, whose projection  for any angle $\theta$ on the noise space is given by $p(\theta)={\mathbf{b}(\theta)}^H\mathbf{U}_{n,d}\mathbf{U}_{n,d}^H{\mathbf{b}(\theta)}$. The 1D search for $\theta$ can be carried out based on\vspace{-0.1cm}
\begin{equation}\label{musictheta}
\hat{\theta}_s=\arg\min_{\theta}{\mathbf{b}(\theta)}^H\mathbf{U}_{n,d}\mathbf{U}_{n,d}^H{\mathbf{b}(\theta)}.\vspace{-0.1cm}
\end{equation}

In the \emph{second stage}, the MLE is adopted for target distance estimation. Assuming the transmitted signal $\mathbf{X}$ is known at the AT, the signals received for target positioning during $T$ time slot follows Gaussian distribution and can be written as\vspace{-0.1cm}
\begin{equation}
    \mathbf{y}_{s,d}=\beta\bm{\delta}(\theta,r)+\mathbf{n}_{s,d},\vspace{-0.1cm}
\end{equation}
where $\bm{\delta}\left(\theta,r\right)=\text{vec}\left(\mathbf{b}(\theta)\mathbf{a}(\theta,r)^T\mathbf{X}\right)$ and $\mathbf{n}_{s,d} \sim \mathcal{CN}(\mathbf{0}, \sigma_d^2\mathbf{I}_{N_bT})$. Let ${\boldsymbol{\eta}}=\left[\theta_s, r_s, \beta_s^r, \beta_s^i\right]^T$ represent the vector that contains all the target parameters, where $\beta_s^r=\Re(\beta_s)$ and $\beta_s^i = \Im(\beta_s)$. With target parameter vector $\bm{\eta}$, the likelihood function of $\mathbf{y}_{s,d}$ is given by $f_{\mathbf{y}_{s,d}}(\mathbf{y}_{s,d};\bm{\eta}) = {(\pi\sigma_d)^{-N_bT/2}}e^{-\sigma_d^{-2}\|\mathbf{y}_{s,d}-\beta\bm{\delta}(\theta,r)\|^2}$. Based on the likelihood function and any given $\theta$ and $r$, $\beta$ can be estimated with MLE, i.e., \vspace{-0.1cm}
\begin{equation}
    \hat{\beta}_s=\arg\min_{\beta}\|\mathbf{y}_{s,d}-\beta\bm{\delta}(\theta,r)\|^2=\frac{\bm{\delta}^H(\theta,r)\mathbf{y}_{s,d}}{\|\bm{\delta}(\theta,r)\|^2}.\vspace{-0.1cm}
\end{equation}
Finally, with $\hat{\beta}$ and $\hat{\theta}$ at hand, the 1D search for $r$ can be carried out based on\vspace{-0.1cm}
\begin{equation}\label{mler}
    \hat{r}_s=\arg\min_{r}\|\mathbf{y}_{s,d}-\hat{\beta}\bm{\delta}(
\hat{\theta}_s,r)\|^2=\arg\max_{r}\frac{|\bm{\delta}^H(\hat{\theta}_s,r)\mathbf{y}_{s,d}|^2}{\|\bm{\delta}(\hat{\theta}_s,r)\|^2}.\vspace{-0.4cm}
\end{equation}}

\noindent The proposed downlink positioning algorithm for target positioning is summarized in \textbf{Algorithm~\ref{algorithm_TS_MUSIC_MLE}}.
\begin{algorithm}[htbp]
\caption{Two-stage downlink positioning algorithm}\label{algorithm_TS_MUSIC_MLE}
\begin{algorithmic}[1]
\STATE {Calculate the approximation of the covariance matrix $\mathbf{R}_{y,d}$ based on the received echo signal $\mathbf{Y}_{s,d}$.}
\STATE {Obtain the noise space $\mathbf{U}_{n,d}$ form $\mathbf{R}_{y,d}$ using the eigenvalue decomposition.}
\STATE \emph{The first stage}: Search for $\hat{\theta}_s$ based on~\eqref{musictheta}.
\STATE \emph{The second stage}: Search for $\hat{r}_s$ based on~\eqref{mler}.
\end{algorithmic}
\end{algorithm}\vspace{-0.5cm}
\subsection{Problem Formulation for Downlink ISPAC}
With estimated target angle and distance $\hat{\theta}_s$ and $\hat{r}_s$, the classic MSE can be used to measure the estimation performance. We adopt the CRB of the estimation of $\theta_s$ and $r_s$ as the performance metric for target positioning since it gives the lower bound of the MSE for user angle and distance and can lead to tractable closed form solutions. 

The FIM of the unknown parameter vector ${\boldsymbol{\eta}}$ during uplink ISPAC can be given by\vspace{-0.1cm}
\begin{equation}\label{eqn:FIM}
    \mathbf{J}_{\boldsymbol{\eta}} = \left[\mathbf{J}_{11}\ \ \mathbf{J}_{12}; \mathbf{J}_{12}^T\ \ \mathbf{J}_{22}\right]\in \mathbb{R}^{4 \times 4},\vspace{-0.1cm}
\end{equation}
where the detailed expressions of ${\bf{J}}_{11}$, ${\bf{J}}_{12}$, and ${\bf{J}}_{22}$ are derived in Appendix~A. Based on \eqref{eqn:FIM}, the CRB for estimating the target angle and distance can be obtained with the inverse formula of the second order matrix as~\cite{kay1993fundamentals}\vspace{-0.1cm}
\begin{equation}\label{eqn:CRB}
    \text{CRB}(\theta_s, r_s) = ({\bf{J}}_{11}-{\bf{J}}_{12}{\bf{J}}_{22}^{-1}{\bf{J}}_{12}^T)^{-1}\in \mathbb{R}^{2 \times 2}.\vspace{-0.1cm}
\end{equation}
During downlink ISPAC, we propose to minimize the trace of the CRB matrix, while guaranteeing the QoS of downlink users. The optimization problem can be formulated as follow:\vspace{-0.1cm}
\begin{subequations}\label{problem:downlink}
    \begin{align}        
        \min_{\mathbf{R}_{d}, \mathbf{W}, \mathbf{F}} \quad &  \text{tr}(\text{CRB}(\theta_s, r_s)) \\[-0.2cm]
        \label{constraint:analog_down}
        \mathrm{s.t.} \quad & \mathbf{F} \in \mathcal{A}_F, \\
        \label{constraint:communication_SINR_down}
        & \text{SINR}_{k,d} \ge \gamma_{k,d}, \forall k \in \mathcal{K},\\
        \label{constraint:sensing1_down}
        & \text{tr}(\mathbf{W}\mathbf{W}^H+\mathbf{R}_{d}) \le P_d,\quad \mathbf{R}_{d}  \succeq  0,
    \end{align}\vspace{-0.5cm}
\end{subequations}

\noindent where constraint \eqref{constraint:analog_down} corresponds to the partially-connected structure of the analog combiner at the receiver. \eqref{constraint:communication_SINR_down} is the minimum SINR constraint of users with $\gamma_{k,d}$ being the minimum SINR of user $k$. \eqref{constraint:sensing1_down} denotes the MT power constraint with $P_d$ being the MT power budget.
\begin{remark}\label{remark2}
Downlink positioning-communication trade-off: \emph{There is a trade-off between positioning and communication performance in the downlink ISPAC framework. A sharing analog precoder is adopted at the MT for sending the probing signal and communication signals. Therefore, The positioning and communication performances should be balanced in the design of the analog transmitting precoding matrix. What’s more, the positioning and communication during downlink ISPAC also share the same power budget. When the communication QoS requirement is stringent, more power needs to be allocated to the communication users, leading to performance degradation for positioning.}  
\end{remark}
To simplify the objective function of \eqref{problem:downlink}, we introduce a positive definite auxiliary matrix $\mathbf{U}$ which satisfies\vspace{-0.1cm}
\begin{equation}\label{eqn:U_J}
{\bf{J}}_{11}-{\bf{J}}_{12}{\bf{J}}_{22}^{-1}{\bf{J}}_{12}^T \succeq \mathbf{U} \succeq 0.\vspace{-0.1cm}
\end{equation}
Based on the monotonicity of traces function over $({\bf{J}}_{11}-{\bf{J}}_{12}{\bf{J}}_{22}^{-1}{\bf{J}}_{12}^T)^{-1}$ and $\mathbf{U}^{-1}$, minimize the trace of the CRB matrix is equivalent to minimize the trace of the $\mathbf{U}^{-1}$ given that \eqref{eqn:U_J} is satisfied. Then problem \eqref{problem:downlink} can be reformulated as follow:\vspace{-0.1cm}
\begin{subequations}\label{problem:downlink_U}
    \begin{align}
        \min_{\mathbf{U}, \mathbf{R}_{d}, \mathbf{W}, \mathbf{F}} \quad &  \text{tr}(\mathbf{U}^{-1}) \\[-0.2cm]
        \label{constraint:FIM_down}
        \mathrm{s.t.} \quad & \left[ {{\bf{J}}_{11}-\mathbf{U}}\ \ {{\bf{J}}_{12}}; {{\bf{J}}_{12}^T}\ \ {{\bf{J}}_{22}}\right] \succeq 0, \\
        \label{constraint:U_down}
        & \mathbf{U} \succeq 0, \\
        & \eqref{constraint:analog_down}-\eqref{constraint:sensing1_down}.
    \end{align}
\end{subequations}\vspace{-0.5cm}

\noindent where the constraints~\eqref{constraint:FIM_down} and~\eqref{constraint:U_down} is equivalent to \eqref{eqn:U_J} according to the Schur complement condition~\cite{zhang2006schur}. To facilitate following derivations, we transform the block-diagonal matrix $\mathbf{F}$ into a diagonal matrix $\tilde{\mathbf{F}}\in \mathbb{C}^{N_a \times N_a}$, which can be given by\vspace{-0.1cm}
\begin{equation}
{\bf{\tilde F}} = \text{Bdiag}\left({\rm{diag}}\left({\bm{f}}_1\right),{\rm{diag}}\left({\bm{f}}_2\right),\cdots,{\rm{diag}}\left({\bm{f}}_{{N_{RF}}}\right)\right),\vspace{-0.1cm}
\end{equation}
$\tilde{\mathbf{F}}$ belongs to a diagonal matrices set $\mathcal{A}_{\tilde{F}}$. Each diagonal element of $\tilde{\mathbf{F}}$, i.e., $[\tilde{\mathbf{F}}]_{nn}$, is constrained by the unit-modulus constraint. The transformation from $\tilde{\mathbf{F}}$ to ${\mathbf{F}}$ can be shown as\vspace{-0.1cm}
\begin{equation}\label{F_transform}
    \mathbf{F}=\tilde{\mathbf{F}}\mathbf{\Phi}=\text{diag}\left(\bm{f}\right)\mathbf{\Phi},\vspace{-0.1cm}
\end{equation}
where vector $\bm{f}=[ {{{\bm{f}}_1}^T,{{\bm{f }}_2}^T, \cdots ,{{\bm{f }}_{{N_{RF}}}^T}} ]^T$, transformation matrix $\mathbf{\Phi}\in \mathbb{C}^{N_a \times N_{RF}}$ is given by\vspace{-0.1cm}
\begin{equation}\label{Phi}
    \mathbf{\Phi} =\left[ {{{\bm{\theta }}_1},{{\bm{\theta }}_2}, \cdots ,{{\bm{\theta }}_{{N_{RF}}}}} \right]= \frac{1}{\sqrt{M}}\text{Bdiag}\left({\bm{1}},{\bm{1}},\cdots,{\bm{1}}\right),\vspace{-0.1cm}
\end{equation}
where $\bm{\theta}_i, \forall i = 1,2,\cdots,N_{RF}$ denotes the $i$-th column of the transformation matrix and
$\mathbf{1}$ is a $M \times 1$ vector with all elements being $1$. To make the FIM constraint \eqref{constraint:FIM_down} easy to handle and remove the coupling between $\mathbf{F}$ and $\tilde{\mathbf{R}}_x$, we define auxiliary matrix $\mathbf{Q}={\mathbf{F}}\tilde{\mathbf{R}}_x{\mathbf{F}}^H=\tilde{\mathbf{F}}\mathbf{\Phi}\tilde{\mathbf{R}}_x\mathbf{\Phi}^H\tilde{\mathbf{F}}^H$. Then the problem \eqref{problem:downlink_U} can be recast as follows:\vspace{-0.1cm}
\begin{subequations}\label{problem:downlink_U_PDD}
    \begin{align}     
        \!\!\!\!\min_{\scriptstyle \mathbf{U},\mathbf{Q},{\mathbf{R}}_{s,d},\atop \scriptstyle\mathbf{W},\tilde{\mathbf{F}}}  &  \text{tr}(\mathbf{U}^{-1}) \\[-0.2cm]
        \label{constraint:PDD_down}
        \mathrm{s.t.} \  & \mathbf{Q} = \tilde{\mathbf{F}}\mathbf{\Phi}\tilde{\mathbf{R}}_x\mathbf{\Phi}^H\tilde{\mathbf{F}}^H, \\
        \label{constraint:FIM_PDD_down}
        & \!\left[ {{\bf{J}}_{11}(\mathbf{Q})\!-\!\mathbf{U}}\ \ {{\bf{J}}_{12}(\mathbf{Q})};{{\bf{J}}_{12}^T(\mathbf{Q})}\ \ {{\bf{J}}_{22}(\mathbf{Q})}\right] \!\succeq\! 0, \\
        \label{constraint:analog_tilde_down}
        & \tilde{\mathbf{F}} \in \mathcal{A}_{\tilde{F}}, \\
        & \eqref{constraint:communication_SINR_down}-\eqref{constraint:sensing1_down}, \eqref{constraint:U_down},
    \end{align}
\end{subequations}\vspace{-0.5cm} 

\noindent where the matrices in the FIM constraint~\eqref{constraint:FIM_PDD_down} are obtained by substituting~\eqref{constraint:PDD_down} into results in Appendix~A. Specifically, they can be expressed as\vspace{-0.1cm}
\begin{equation}
    \!\mathbf{J}_{11}(\mathbf{Q})\!\! =\!\!\frac{2 |\beta|^2 T }{\sigma_d^2}   \Re \!\left\{\!    \begin{bmatrix}
         \mathrm{tr}(\dot{\mathbf{G}}_{\theta_{s}}^T \mathbf{Q} \dot{\mathbf{G}}_{\theta_{s}}^*) \!\!\!&\mathrm{tr}( \dot{\mathbf{G}}_{\theta_{s}}^T \mathbf{Q} \dot{\mathbf{G}}_{r_s}^* )\\
         \mathrm{tr}(\dot{\mathbf{G}}_{r_s}^T \mathbf{Q} \dot{\mathbf{G}}_{\theta_{s}}^* )\!\!\!&  \mathrm{tr}(\dot{\mathbf{G}}_{r_s}^T \mathbf{Q} \dot{\mathbf{G}}_{r_s}^*)
    \end{bmatrix}\!\right\}.
\end{equation}
\begin{equation}
    \mathbf{J}_{12}(\mathbf{Q}) = \frac{2 T}{\sigma_d^2} \Re \left( \begin{bmatrix}
        \beta^* \mathrm{tr}( \tilde{\mathbf{G}}^T \mathbf{Q}  \dot{\mathbf{G}}_{\theta_s}^* ) \\
        \beta^* \mathrm{tr}( \tilde{\mathbf{G}}^T \mathbf{Q}  \dot{\mathbf{G}}_{r_s}^* )
    \end{bmatrix} [1, j] \right),
    \end{equation}
\begin{equation}
    \mathbf{J}_{22}(\mathbf{Q}) = \frac{2 T}{\sigma_d^2} \mathbf{I}_2 \mathrm{tr} ( \tilde{\mathbf{G}}^T \mathbf{Q} \tilde{\mathbf{G}}^* ).\vspace{-0.2cm}
\end{equation}
Problem \eqref{problem:downlink_U_PDD} is non-convex due to the non-convex SINR constraint \eqref{constraint:communication_SINR_down}, the quadratic constraint \eqref{constraint:PDD_down}, and the unit modulus constraint \eqref{constraint:analog_tilde_down}. We adopt the PDD optimization technique to solve this problem~\cite{9120361}. Specifically, in the outer iteration loop, the Lagrangian dual matrix and penalty factor are updated with rules given in~\cite{9120361}. In the inner iteration loop, the AL problem is solved. By introducing the Lagrangian dual matrix ${\bm{\Upsilon}}$ and the penalty factor $\rho$ for constraint \eqref{constraint:PDD_down}, the augmented Lagrangian (AL) problem can be given as:\vspace{-0.1cm}
\begin{subequations}\label{problem:downlink_PDD}
    \begin{align}        
       \!\! \min_{\scriptstyle \mathbf{U},\mathbf{Q},{\mathbf{R}}_{s,d},\atop \scriptstyle\mathbf{W},\tilde{\mathbf{F}}} \  &  \text{tr}(\mathbf{U}^{-1})+\frac{1}{2\rho}\|\mathbf{Q}-\tilde{\mathbf{F}}\mathbf{\Phi}\tilde{\mathbf{R}}_x\mathbf{\Phi}^H\tilde{\mathbf{F}}^H-\rho{\bm{\Upsilon}} \|^2 \\[-0.2cm]
        \mathrm{s.t.} \  & \eqref{constraint:communication_SINR_down}-\eqref{constraint:sensing1_down}, \eqref{constraint:U_down}, \eqref{constraint:FIM_PDD_down}, \eqref{constraint:analog_tilde_down}.
    \end{align}
\end{subequations}\vspace{-0.5cm}

\noindent For a given dual matrix and penalty factor, it can be observed that constraints of the AL problem are separable. Thus, the AO method is adopted to address it with variables in two groups $\{\mathbf{U},\mathbf{Q},{\mathbf{R}}_s,\mathbf{W}\}$ and $\{\mathbf{F}\}$, i.e., the variables of one group is optimized with variables in the other group fixed in every iteration, and this leads to the following two subproblems in each AO iteration. \vspace{-0.2cm}
\subsection{Subproblem with respect to $\{\mathbf{U},\mathbf{Q},{\mathbf{R}}_{s,d},\mathbf{W}\}$} 

The SINR constraint for user $k$ in \eqref{constraint:communication_SINR_down} can be rewritten as \vspace{-0.1cm}
\begin{equation}\label{downlink_PDD_RW_SDR_SINR1}
    \frac{1}{\gamma_{k,d}}|\tilde{\mathbf{h}}_k^T{\mathbf{w}}_{k,d}|^2 \ge \sum\nolimits_{i\ne k}|\tilde{\mathbf{h}}_k^T{\mathbf{w}}_{i,d}|^2+\tilde{\mathbf{h}}_k^T{\mathbf{R}}_{s,d}\tilde{\mathbf{h}}_k^*+\sigma_0^2. \vspace{-0.1cm}
\end{equation}
where $\tilde{\mathbf{h}}_k=\mathbf{F}^T{\mathbf{h}}_k$. Recall that $\tilde{\mathbf{R}}_x=\mathbf{W}\mathbf{W}^H+\mathbf{R}_{d}$ in \eqref{RX} and define $\mathbf{W}_{k,d}=\mathbf{w}_{k,d}\mathbf{w}_{k,d}^H$. The subproblem with respect to $\{\mathbf{U},\mathbf{Q},{\mathbf{R}}_{s,d},\mathbf{W}\}$ can be formulated as: \vspace{-0.1cm}
\begin{subequations}\label{problem:downlink_PDD_RW_SDR}
    \begin{align}       
        \min_{\scriptstyle \mathbf{U},\mathbf{Q},\tilde{\mathbf{R}}_x,\atop \scriptstyle\mathbf{W}_{k,d}} \  &  \text{tr}(\mathbf{U}^{-1})+\frac{1}{2\rho}\|\mathbf{Q}-\tilde{\mathbf{F}}\mathbf{\Phi}\tilde{\mathbf{R}}_x\mathbf{\Phi}^H\tilde{\mathbf{F}}^H-\rho{\bm{\Upsilon}} \|^2 \\[-0.2cm]
        \label{downlink_PDD_RW_SDR_SINR}
        \mathrm{s.t.} \  & (1+\frac{1}{\gamma_{k,d}})\tilde{\mathbf{h}}_k^T{\mathbf{W}}_{k,d}\tilde{\mathbf{h}}_k^* \ge \tilde{\mathbf{h}}_k^T\tilde{\mathbf{R}}_x\tilde{\mathbf{h}}_k^*+\sigma_0^2, \forall k \\      
        & \mathbf{W}_{k,d} \succeq 0, \text{rank}(\mathbf{W}_{k,d}) = 1, \forall k, \\
        & \text{tr}(\tilde{\mathbf{R}}_x) \le P, \tilde{\mathbf{R}}_x \succeq \sum\nolimits_k\mathbf{W}_{k,d}, \eqref{constraint:U_down}, \eqref{constraint:FIM_PDD_down},
    \end{align}
\end{subequations}\vspace{-0.5cm}

\noindent where the constraint~\eqref{downlink_PDD_RW_SDR_SINR} is transformed from the communication SINR constraint~\eqref{downlink_PDD_RW_SDR_SINR1}. The SDR problem of problem~\eqref{problem:downlink_PDD_RW_SDR} is convex and its optimal solution, i.e., $\{\mathbf{W}_{k,d}^\text{SDR}, \forall k\}$, can be solved via CVX~\cite{grant2014cvx}. Due to the SDR, matrices $\{\mathbf{W}_{k,d}^\text{SDR}, \forall k\}$ have general rank. \textbf{Theorem $1$} in~\cite{9124713} can be used to constructed optimal solutions to problem~\eqref{problem:downlink_PDD_RW_SDR} using $\{\mathbf{W}_{k,d}^\text{SDR}, \forall k\}$. Specifically, suppose $\{\mathbf{W}_{k,d}^\text{SDR}, \forall k\}$ are the optimal solution to the SDR problem of problem~\eqref{problem:downlink_PDD_RW_SDR}, the optimal solution to the original problem~\eqref{problem:downlink_PDD_RW_SDR} is given by\vspace{-0.1cm}
\begin{equation}
    \mathbf{W}_{k,d}^\star\!=\!(\tilde{\mathbf{h}}_k^T\mathbf{W}_{k,d}^\text{SDR}\tilde{\mathbf{h}}_k^*)^{-\frac{1}{2}}\mathbf{W}_{k,d}^\text{SDR}\tilde{\mathbf{h}}_k^*\tilde{\mathbf{h}}_k^T\mathbf{W}_{k,d}^\text{SDR}(\tilde{\mathbf{h}}_k^T\mathbf{W}_{k,d}^\text{SDR}\tilde{\mathbf{h}}_k^*)^{-\frac{1}{2}}.\vspace{-0.2cm}
\end{equation}

\subsection{Subproblem with respect to $\{\mathbf{F}\}$} The subproblem with respect to $\{\mathbf{F}\}$ can be expressed as the subproblem with respect to $\{\tilde{\mathbf{F}}\}$, which can be given as follows:\vspace{-0.2cm}
\begin{subequations} \label{problem:downlink_PDD_F}  
    \begin{align}
        \min_{\tilde{\mathbf{F}}} \quad &  \|\mathbf{Q}-\tilde{\mathbf{F}}\mathbf{\Phi}\tilde{\mathbf{R}}_x\mathbf{\Phi}^H\tilde{\mathbf{F}}^H-\rho{\bm{\Upsilon}} \|^2 \\[-0.2cm]
        \mathrm{s.t.} \quad & \eqref{constraint:communication_SINR_down}, \eqref{constraint:analog_tilde_down}.
    \end{align}
\end{subequations}\vspace{-0.5cm}

\noindent The eigenvalue decomposition of $\mathbf{\Phi}\tilde{\mathbf{R}}_x\mathbf{\Phi}^H$ with a rank of $R$ is \vspace{-0.1cm}
\begin{equation}
\mathbf{\Phi}\tilde{\mathbf{R}}_x\mathbf{\Phi}^H = \sum\nolimits_{r=1}^{R}\rho_r\mathbf{v}_r\mathbf{v}_r^H,\vspace{-0.1cm}
\end{equation}
where $\rho_r$ and $\mathbf{v}_r$ are the eigenvalue and the associated eigenvector of $\mathbf{\Phi}\tilde{\mathbf{R}}_x\mathbf{\Phi}^H$. To facilitate the following derivations, define $\hat{\mathbf{v}}_r=\sqrt{\rho_r}{\mathbf{v}}_r$, ${\mathbf{V}}_r=\hat{\mathbf{v}}_r\hat{\mathbf{v}}_r^H$, and $\dot{\mathbf{F}}=\bm{f}\bm{f}^H$. Then we have\vspace{-0.1cm}
\begin{equation}
\tilde{\mathbf{F}}\mathbf{\Phi}\tilde{\mathbf{R}}_x\mathbf{\Phi}^H\tilde{\mathbf{F}}^H=\tilde{\mathbf{F}}\sum\nolimits_{r=1}^{R}\hat{\mathbf{v}}_r\hat{\mathbf{v}}_r^H\dot{\mathbf{F}}^H=\sum\nolimits_{r=1}^{R}\mathbf{V}_r\hat{\mathbf{F}}\mathbf{V}_r^H.\vspace{-0.1cm}
\end{equation}
Problem \eqref{problem:downlink_PDD_F} can be reformulated as \vspace{-0.1cm}
\begin{subequations}
    \label{problem:downlink_PDD_F_SDR}
    \begin{align}
        \min_{\dot{\mathbf{F}}} \quad &  \|\mathbf{Q}-\sum\nolimits_{r=1}^{R}\mathbf{V}_r\dot{\mathbf{F}}\mathbf{V}_r^H-\rho{\bm{\Upsilon}} \|^2 \\[-0.2cm]
        \label{constraint:SINR_F_downlink}
        \mathrm{s.t.} \quad & \frac{1}{\gamma_{k,d}}\text{tr}(\dot{\mathbf{F}}\mathbf{H}_k^k) \le \text{tr}(\dot{\mathbf{F}}(\sum\nolimits_{i\ne k}\mathbf{H}_k^i+\mathbf{{\bf{\Psi }}}_k))+\sigma_0^2, \\
        \label{constraint:modu_F_downlink}
        & [\dot{\mathbf{F}}]_{n,n} =1 , \forall n = 1, 2, \cdots, N_a, \\
        \label{constraint:SEMIdefinit_F_downlink}
        & \dot{\mathbf{F}} \succeq 0, \text{rank}(\dot{\mathbf{F}}) =1,
    \end{align}
\end{subequations}\vspace{-0.5cm}

\noindent where $\mathbf{H}_k^i=\text{diag}\left(\mathbf{h}_k\right)\mathbf{\Phi}\mathbf{w}_{i,d}$ and $\mathbf{{\bf{\Psi }}}_k=\text{diag}\left(\mathbf{h}_k\right)\mathbf{\Phi}\mathbf{R}_{d}\mathbf{\Phi}^H\text{diag}\left(\mathbf{h}_k^H\right)$. The SDR problem of problem \eqref{problem:downlink_PDD_F_SDR} is convex and its optimal solution can be solved via CVX. The solutions of $\dot{\mathbf{F}}$ that satisfy the rank-one constraint can be obtained by using Gaussian randomization~\cite{5447068}.
\subsection{The Overall Algorithm for Solving~\eqref{problem:downlink_U_PDD}}
The proposed algorithm for solving~\eqref{problem:downlink_U_PDD} during the downlink ISPAC is summarized in \textbf{Algorithm~\ref{algorithm4}}. Every iteration of \textbf{Algorithm~\ref{algorithm4}} yields a non-increasing objective function value for~\eqref{problem:downlink_U_PDD}. Furthermore, because the BS power budget is constrained, the objective function value of~\eqref{problem:downlink_U_PDD} is lower bounded. Thus, \textbf{Algorithm~\ref{algorithm4}} is certain to converge to a stationary point of~\eqref{problem:downlink_U_PDD}. Given the solution accuracy $\epsilon_{\ref{algorithm4},\text{AO}}$ and $\epsilon_{\ref{algorithm4},\text{PDD}}$, the complexity of solving problem~\eqref{problem:downlink_U_PDD} adopting the interior-point method is $\mathcal{O}(\log(1/\epsilon_{\ref{algorithm4},\text{PDD}})\log(1/\epsilon_{\ref{algorithm4},\text{AO}})(N_{RF}^{4.5}+N_a^{4.5}))$~\cite{5447068}.\vspace{-0.2cm}
\begin{algorithm}[htbp]
\caption{PDD-based algorithm for solving problem~\eqref{problem:downlink_U_PDD}}\label{algorithm4}
\begin{algorithmic}[1]
\STATE {Initialize feasible $\mathbf{F}, {\bm{\Upsilon}}^{(n)}$, $\eta^{(n)}$, and ${\rho}^{(n)}$ with $n=0$, set $\mu \in (0,1)$.}
\STATE {\bf repeat: outer loop}
\STATE \quad Obtain AL problem ~\eqref{problem:downlink_PDD} with respect to ${\bm{\Upsilon}}^{(n)}$ and~${\rho}^{(n)}$.\STATE \quad {\bf repeat: inner loop}
\STATE \quad\quad Given $\mathbf{F}=\tilde{\mathbf{F}}\mathbf{\Phi}$, obtain $\mathbf{Q}, \mathbf{R}_{d}, \mathbf{W}$ by solving problem~\eqref{problem:downlink_PDD_RW_SDR}.
\STATE \quad\quad Given $\mathbf{Q}, \mathbf{R}_{d}, \mathbf{W}$, obtain $\mathbf{F}=\tilde{\mathbf{F}}\mathbf{\Phi}$ by solving problem~\eqref{problem:downlink_PDD_F}.
\STATE \quad {\bf until} the fractional decrease of~\eqref{problem:downlink_PDD} is less than threshold $\epsilon_{\ref{algorithm4},\text{AO}}$. 
\STATE \quad Obtain $\mathbf{F}^{(n+1)}=\tilde{\mathbf{F}}^{(n+1)}\mathbf{\Phi}$, $\mathbf{Q}^{(n+1)}$, and $\tilde{\mathbf{R}}_x^{(n+1)}=\mathbf{R}_{d}^{(n+1)}+\mathbf{W}^{(n+1)}(\mathbf{W}^{(n+1)})^H$.
\STATE \quad {\bf if} $\|\mathbf{Q}^{(n+1)}-\tilde{\mathbf{F}}^{(n+1)}\mathbf{\Phi}\tilde{\mathbf{R}}_x^{(n+1)}\mathbf{\Phi}^H(\tilde{\mathbf{F}}^{(n+1)})^H\|_\infty \le \eta^{(n)}$ {\bf then}
\STATE \quad \quad Update ${\bm{\Upsilon}}$ with ${\bm{\Upsilon}}^{(n+1)}={\bm{\Upsilon}}^{(n)}+\frac{1}{\rho^{(n)}}(\mathbf{Q}^{(n+1)}-\tilde{\mathbf{F}}^{(n+1)}\mathbf{\Phi}\tilde{\mathbf{R}}_x^{(n+1)}\mathbf{\Phi}^H(\tilde{\mathbf{F}}^{(n+1)})^H)$.
\STATE \quad \quad Penalty factor remains the same, i.e., $\rho^{(n+1)}=\rho^{(n)}$.\STATE \quad {\bf else}
\STATE \quad \quad Update $\rho$ with $\rho^{(n+1)}=\mu\rho^{(n)}$.
\STATE \quad \quad Lagrangian dual matrix ${\bm{\Upsilon}}$ remains the same, i.e., ${\bm{\Upsilon}}^{(n+1)}={\bm{\Upsilon}}^{(n)}$.
\STATE \quad {\bf end if}
\STATE \quad Set $\eta^{(n+1)}=0.9\|\mathbf{Q}^{(n+1)}-\tilde{\mathbf{F}}^{(n+1)}\mathbf{\Phi}\tilde{\mathbf{R}}_x^{(n+1)}\mathbf{\Phi}^H(\tilde{\mathbf{F}}^{(n+1)})^H\|_\infty$, $n=n+1$.
\STATE {\bf until} the constraint violation is less than threshold $\epsilon_{\ref{algorithm4},\text{PDD}}$.
\end{algorithmic}
\end{algorithm}\vspace{-0.4cm}

\section{Target Positioning and CRB Optimization Design for Uplink ISPAC} 

In this section, we propose the two-stage uplink positioning algorithm for estimating the target location. Then, we design the uplink ISPAC system based on the CRB.\vspace{-0.2cm}
\subsection{Positioning Algorithm for Uplink ISPAC}
In terms of uplink target positioning, our purpose is to obtain angle $\theta_s$ and distance $r_s$ of the sensing target utilizing the acquired echo signals $\mathbf{Y}_{s,u}$. Inspired by the parameter split technique used in~\cite{8359308}, a customised uplink positioning algorithm is proposed to cut down the computational cost, where the angle and distance of the target are obtained with two successive 1D MUSIC searches.

Specifically, the covariance matrix of the received echo signal can be approximated as $\mathbf{R}_{y,u}=\frac{1}{T}\mathbf{Y}_{s,u}\mathbf{Y}_{s,u}^H$. Using the eigenvalue decomposition, the noise space $\mathbf{U}_{n,u}$ with the dimension of $N_{RF}-1$ is obtained. Based on~\eqref{eqn:ys}, the received probing signal after analog combining is spanned by the vector $\tilde{\mathbf{a}}(\theta,r)=\mathbf{F}^H\mathbf{a}(\theta,r)$. According to the structure of the near-field array response vector, we split it as \vspace{-0.1cm}
\begin{equation}\label{split}
\begin{aligned}
\tilde{\mathbf{a}}&(\theta,r) =\mathbf{F}^H\text{diag}(1, e^{j\vartheta}, \cdots, e^{j(N_a-1)\vartheta})\times \\ &\left[1, e^{j\varphi}, \cdots, e^{j(N_a-1)^2\varphi}\right]^H= \mathbf{F}^H\text{diag}\left(\mathbf{c}(\vartheta)\right){\mathbf{d}(\varphi)},
\end{aligned} \vspace{-0.1cm}
\end{equation}
where $\vartheta=k_cd\sin\theta$ and $\varphi=\frac{k_cd^2\cos^2\theta}{2r}$. Then the projection of $\tilde{\mathbf{a}}(\theta,r)$ for any angle $\theta$ and distance $r$ on the noise space is given by \vspace{-0.1cm}
\begin{equation}
p(\theta,r)={\mathbf{d}(\varphi)}^H\mathbf{\Gamma}(\vartheta){\mathbf{d}(\varphi)}, \vspace{-0.1cm}
\end{equation}
where $\mathbf{\Gamma}(\vartheta)=\text{diag}\left(\mathbf{c}(\vartheta)\right)^H\mathbf{F}\mathbf{U}_{n,u}\mathbf{U}_{n,u}^H\mathbf{F}^H\text{diag}\left(\mathbf{c}(\vartheta)\right)$. With above reformulation, the 2D angle and distance search can be expressed as following optimization problem \vspace{-0.1cm}
\begin{subequations}
    \label{MUSIC}
    \begin{align}
        \min_{\vartheta, \varphi} \quad &  {\mathbf{d}(\varphi)}^H\mathbf{\Gamma}(\vartheta){\mathbf{d}(\varphi)} \\[-0.1cm]
        \label{music_b}
        & \mathbf{e}_1^T{\mathbf{d}(\varphi)}=1,
    \end{align}
\end{subequations}\vspace{-0.5cm}

\noindent where $\mathbf{e}_1^T=[1,0,\cdots,0]\in\mathbb{R}^{1\times N_a}$. In the first stage, we focus on searching for $\vartheta$ and define the Lagrangian function of problem~\eqref{MUSIC} as \vspace{-0.1cm}
\begin{equation}
\mathcal{L}(\vartheta, \varphi, \tau)={\mathbf{d}(\varphi)}^H\mathbf{\Gamma}(\vartheta){\mathbf{d}(\varphi)}-\tau(\mathbf{e}_1^T{\mathbf{d}(\varphi)}-1),\vspace{-0.1cm}
\end{equation}
where $\tau\ge0$ is the Lagrange multiplier. Exploiting the Karush-Kuhn-Tucker (KKT) conditions of stationarity point, i.e., $\frac{\partial\mathcal{L}}{\partial\mathbf{d}}=2\mathbf{\Gamma}(\vartheta){\mathbf{d}(\varphi)}+\tau\mathbf{e}_{1}=0$ and considering constraint~\eqref{music_b}, we have \vspace{-0.1cm}
\begin{equation}
\mathbf{d}^\text{opt}=\frac{\mathbf{\Gamma}(\vartheta)^{-1}\mathbf{e}_{1}}{\mathbf{e}_{1}^H\mathbf{\Gamma}(\vartheta)^{-1}\mathbf{e}_{1}}.\vspace{-0.1cm}
\end{equation}
In the \emph{first stage}, the 1D search for $\vartheta$ can be carried out based on\vspace{-0.1cm}
\begin{equation}\label{forvartheta}
\!\hat{\vartheta}\!=\!\arg\min_{\vartheta}(\mathbf{d}^\text{opt})^H\mathbf{\Gamma}(\vartheta)\mathbf{d}^\text{opt}\!=\!\arg\max_{\vartheta}\mathbf{e}_{1}^H(\mathbf{\Gamma}(\vartheta))^{-1}\mathbf{e}_{1}.\!\vspace{-0.1cm}
\end{equation}
In the \emph{second stage}, with the search result $\hat{\vartheta}$ for target angle, the 1D search for $\varphi$ can be carried out by solving\vspace{-0.1cm}
\begin{equation}\label{varphi}
\hat{\varphi}=\arg\min_{\varphi}\mathbf{d}(\varphi)^H\mathbf{\Gamma}(\hat{\vartheta})\mathbf{d}(\varphi).\vspace{-0.1cm}
\end{equation}
Finally, the transformation between searching result $\{\hat{\vartheta},\hat{\varphi}\}$ and target angle and distance is given by\vspace{-0.1cm}
\begin{equation}\label{transformation}
\hat{\theta}_s=\arcsin\frac{\hat{\vartheta}}{k_cd},\quad \hat{r}_s=\frac{k_cd^2\cos^2\hat{\theta}_s}{2\hat{\varphi}}.\vspace{-0.1cm}
\end{equation}
The proposed uplink positioning algorithm for target positioning is summarized in \textbf{Algorithm~\ref{algorithm_TS_MUSIC}}.
\begin{algorithm}[htbp]
\caption{Two-stage uplink positioning algorithm}\label{algorithm_TS_MUSIC}
\begin{algorithmic}[1]
\STATE {Calculate the approximation of the covariance matrix $\mathbf{R}_{y,u}$ based on the received echo signal $\mathbf{Y}_{s,u}$.}
\STATE {Obtain the noise space $\mathbf{U}_{n,u}$ form $\mathbf{R}_{y,u}$ using the eigenvalue decomposition.}
\STATE \emph{The first stage}: Search for $\hat{\vartheta}$ based on~\eqref{forvartheta}.
\STATE \emph{The second stage}: Search for $\hat{\varphi}$ based on~\eqref{varphi}.
\STATE Carry out the transformation~\eqref{transformation} to acquire $\hat{\theta}_s$ and $\hat{r}_s$.
\end{algorithmic}
\end{algorithm}\vspace{-0.2cm}
\subsection{Problem Formulation for Uplink ISPAC}
The FIM of the unknown target parameter vector ${\boldsymbol{\eta}}$ can be given as the form in~\eqref{eqn:FIM}. The detailed expressions of ${\bf{J}}_{11}$, ${\bf{J}}_{12}$, and ${\bf{J}}_{22}$ in downlink ISPAC are derived in Appendix~B. The CRB for estimating the target angle and distance can be given by\vspace{-0.1cm}
\begin{equation}\label{eqn:CRB_down}
    \text{CRB}(\theta_s, r_s) = ({\bf{J}}_{11}-{\bf{J}}_{12}{\bf{J}}_{22}^{-1}{\bf{J}}_{12}^T)^{-1}\in \mathbb{R}^{2 \times 2}.\vspace{-0.1cm}
\end{equation}
We propose to minimize the trace of the CRB matrix while guaranteeing the QoS of uplink users. The optimization problem can be formulated as follow:\vspace{-0.1cm}
\begin{subequations}\label{problem:uplink}
    \begin{align}        
        \min_{\mathbf{R}_{u}, \mathbf{w}_{k,u}, \mathbf{F}} \quad &  \text{tr}(\text{CRB}(\theta_s, r_s)) \\[-0.2cm]
        \label{constraint:analog}
        \mathrm{s.t.} \quad & \mathbf{F} \in \mathcal{A}_F, \\
        \label{constraint:communication_SINR}
        & \text{SINR}_{k,u} \ge \gamma_{k,u}, \forall k \in \mathcal{K},\\        
        \label{constraint:sensing1}
        & \text{tr}(\mathbf{R}_{u}) \le P_s,\quad \mathbf{R}_{u}  \succeq  0,
    \end{align}\vspace{-0.5cm}
\end{subequations}

\noindent where constraint \eqref{constraint:analog} corresponds to the partially-connected structure of the analog combiner at the receiver. \eqref{constraint:communication_SINR} is the minimum SINR constraint of users with $\gamma_{k,u}$ being the lower bound for the SINR of user $k$. \eqref{constraint:sensing1} denotes the positioning power constraint with $P_s$ being the power budget at the AT.
\begin{remark}\label{remark1}
Uplink positioning-communication trade-off: \emph{There is a trade-off between positioning and communication performance in the uplink ISPAC framework. On the one hand, the echoed probing signal causes positioning to communication (P2C) interference, resulting in the degradation of uplink communication performance; On the other hand, the MT simultaneously gathers the echo probing signal and communication signals with a sharing analog combiner, thus the design of the analog combining matrix at the MT should strike a balance between positioning and communication performances.}   
\end{remark}
To Simplify the objective function of~\eqref{problem:uplink}, introduce a positive definite auxiliary matrix $\mathbf{U}$ which satisfies\vspace{-0.1cm}
\begin{equation}
{\bf{J}}_{11}-{\bf{J}}_{12}{\bf{J}}_{22}^{-1}{\bf{J}}_{12}^T \succeq \mathbf{U} \succeq 0.\vspace{-0.1cm}
\end{equation}
Then problem~\eqref{problem:uplink} can be reformulated as follow:\vspace{-0.1cm}
\begin{subequations}\label{problem:uplink_U}
    \begin{align}
        \min_{\mathbf{U}, \mathbf{R}_{u}, \mathbf{F}, \mathbf{w}_{k,u}} \quad &  \text{tr}(\mathbf{U}^{-1}) \\[-0.2cm]
        \label{constraint:FIM}
\mathrm{s.t.} \quad & \left[{{\bf{J}}_{11}-\mathbf{U}}\ \ {{\bf{J}}_{12}};{{\bf{J}}_{12}^T}\ \ {{\bf{J}}_{22}}\right] \succeq 0, \\
        \label{constraint:U}
        & \mathbf{U} \succeq 0, \\
        & \eqref{constraint:analog}-\eqref{constraint:sensing1},
    \end{align}
\end{subequations}\vspace{-0.5cm}

\noindent In the following, we address the optimization problem \eqref{problem:uplink_U} with the AO method. Specifically, we partition the optimization variables into two blocks, i.e., $\{{\mathbf{R}}_{s,u},\mathbf{w}_{k,u}\}$ and $\{\mathbf{F}\}$. During each iteration, we successively optimize the variables within one block, while keeping the variables in the other block constant.\vspace{-0.2cm}
\subsection{Subproblem with respect to $\{\mathbf{U},{\mathbf{R}}_{s,u},\mathbf{w}_{k,u}\}$:} The subproblem with respect to $\{\mathbf{U},{\mathbf{R}}_{s,u},\mathbf{w}_{k,u}\}$ is giving by\vspace{-0.1cm}
\begin{subequations}\label{problem:uplink_RW}
    \begin{align}       
        \min_{\mathbf{U}, \mathbf{R}_{u}, \mathbf{w}_{k,u}} \    &  \text{tr}(\mathbf{U}^{-1}) \\[-0.2cm]
        \label{constraint:FIM_RW}
        \mathrm{s.t.} \   & \left[{\bf{J}}_{11}(\mathbf{R}_{u})-\mathbf{U}\ {{\bf{J}}_{12}(\mathbf{R}_{u})};
{{\bf{J}}_{12}^T(\mathbf{R}_{u})}\ {{\bf{J}}_{22}(\mathbf{R}_{u})}
 \right] \succeq 0, \\
        & \eqref{constraint:communication_SINR}-\eqref{constraint:sensing1}, \eqref{constraint:U},
    \end{align}
\end{subequations}\vspace{-0.5cm}

\noindent where the entries of the matrices in \eqref{constraint:FIM_RW} are given as in Appendix~B. The optimization problem is non-convex due to the non-convex user SINR constraint \eqref{constraint:communication_SINR}, which can be reformulated as  in~\eqref{constraint:communication_SINR_REFOR} at the top of next page,
\begin{figure*}[!t]
\normalsize\vspace{-0.3cm}
\begin{equation}\label{constraint:communication_SINR_REFOR}
\mathbf{w}_{k,u}^H \hat{\mathbf{G}} \mathbf{R}_{u} \hat{\mathbf{G}}^H  \mathbf{w}_{k,u} + \sum\nolimits_{i\neq k} |\mathbf{w}_{k,u}^H \hat{\mathbf{h}}_{i}|^2 - \frac{|\mathbf{w}_{k,u}^H  \hat{\mathbf{h}}_{k}|^2}{\gamma_{k,u}} + \sigma_u^2 \|\mathbf{w}_{k,u}\|^2 = \Pi_k + \text{tr} (\mathbf{W}_{k,u}\mathbf{A}_{k}) \le 0.\vspace{-0.2cm}
\end{equation}
\vspace{-0.2cm}
\end{figure*}
where $\Pi_k = 2\text{tr} (\mathbf{W}_{k,u}\hat{\mathbf{G}} \mathbf{R}_{u} \hat{\mathbf{G}}^H)$, $\hat{\mathbf{h}}_{k}=\sqrt{P_u}\mathbf{F}^H \mathbf{h}_{k}$,  $\hat{\mathbf{G}}=\mathbf{F}^H \mathbf{G}$, $\mathbf{W}_{k,u}=\mathbf{w}_{k,u}\mathbf{w}_{k,u}^H$, and $\mathbf{A}_k=\sum_{i\neq k}\hat{\mathbf{h}}_i\hat{\mathbf{h}}_i^H-\frac{1}{\gamma_{k,u}}\hat{\mathbf{h}}_k\hat{\mathbf{h}}_k^H+\sigma_u^2 \mathbf{I}_{N_{RF}}, \forall k \in \mathcal{K}$. Term $\Pi_k$ can be rewritten as:\vspace{-0.1cm}
\begin{equation}
    \!\Pi_k  \!=\! \|\mathbf{W}_{k,u}+\hat{\mathbf{G}} \mathbf{R}_{u} \hat{\mathbf{G}}^H\|_{F}^2-\|\mathbf{W}_{k,u}\|_{F}^2-\|\hat{\mathbf{G}} \mathbf{R}_{u} \hat{\mathbf{G}}^H\|_{F}^2,\vspace{-0.1cm}
\end{equation}
Then, the SCA technique is adopted and the non-convex term $\Pi_k$ is replaced by its upper bound $\Pi_k^{up}$. Specifically, for a given point $\{\mathbf{W}_{k,u}^{(n)}, \mathbf{R}_{u}^{(n)}\}$ in the $n$-th SCA iteration, using the first-order Taylor expansion, the convex upper bound of $\Pi_k$ can be expressed as in~\eqref{gamma Bm CDF} at the top of next page.
\begin{figure*}[!t]
\normalsize\vspace{-0.3cm}
\begin{equation}\label{gamma Bm CDF}
\Pi_k^{up} = \|\mathbf{W}_{k,u}+\hat{\mathbf{G}} \mathbf{R}_{u} \hat{\mathbf{G}}^H\|_{F}^2+\|\mathbf{W}_{k,u}^{(n)}\|_{F}^2-\text{tr} ((\mathbf{W}_{k,u}^{(n)})^H\mathbf{W}_{k,u})+\|\hat{\mathbf{G}} \mathbf{R}_{u}^{(n)} \hat{\mathbf{G}}^H\|_{F}^2 -\text{tr} ((\hat{\mathbf{G}}^H\hat{\mathbf{G}} \mathbf{R}_{u}^{(n)} \hat{\mathbf{G}}^H\hat{\mathbf{G}})^H\mathbf{R}_{u}).\vspace{-0.2cm}
\end{equation}
\hrulefill \vspace*{0pt}\vspace{-0.3cm}
\end{figure*}
For the $n$-th SCA iteration, problem \eqref{problem:uplink_RW} can be transformed into following problem:\vspace{-0.1cm}
\begin{subequations}\label{problem:uplink_RW_SCA}
    \begin{align}        
        \min_{\mathbf{U}, \mathbf{R}_{u}, \mathbf{w}_{k,u}} \quad &  \text{tr}(\mathbf{U}^{-1}) \\[-0.2cm]
        \mathrm{s.t.} \quad \label{constraint:communication_SINR_SCA}
        & \text{tr} \left(\mathbf{W}_{k,u}\mathbf{A}_{k}\right) + \Pi_k^{up} \le 0, \forall k \in \mathcal{K},\\
        \label{constraint:rank_W}
        & \mathbf{W}_{k,u} \succeq 0, \text{rank}(\mathbf{W}_{k,u})=1, \forall k \in \mathcal{K},\\
        & \eqref{constraint:sensing1}, \eqref{constraint:U}, \eqref{constraint:FIM_RW},
    \end{align}
\end{subequations}\vspace{-0.5cm}

\noindent where constraint~\eqref{constraint:rank_W} comes from the definition of $\mathbf{W}_{k,u}, \forall k \in \mathcal{K}$. The SDR problem of problem \eqref{problem:uplink_RW_SCA} is convex and its optimal solution can be solved via CVX. Although the SDR may result in solutions with general rank, rank-one solutions can be obtained by using Gaussian randomization. The proposed SCA algorithm for solving subproblem~\eqref{problem:uplink_RW} is summarized in \textbf{Algorithm~\ref{algorithm1}}. 

\begin{algorithm}[htbp]
\caption{SCA algorithm for solving problem~\eqref{problem:uplink_RW}}\label{algorithm1}
\begin{algorithmic}[1]
\STATE {Initialize feasible point $\mathbf{R}_{u}^{(0)}$ and $\mathbf{W}_{k,u}^{(0)}, \forall k \in \mathcal{K}$ and set $n=0$.}
\STATE {\bf repeat:}
\STATE \quad Get intermediate solutions by solving~\eqref{problem:uplink_RW_SCA} for given $\mathbf{R}_{u}^{(n)}$ and $\mathbf{W}_{k,u}^{(n)}, \forall k \in \mathcal{K}$.
\STATE \quad Update $\mathbf{R}_{u}^{(n+1)}$ and $\mathbf{W}_{k,u}^{(n+1)}, \forall k \in \mathcal{K}$ with the intermediate solutions and set $n=n+1$.
\STATE {\bf until} the fractional decrease of~\eqref{problem:uplink_RW} is less than threshold~$\epsilon_{\ref{algorithm1},\text{SCA}}$.
\end{algorithmic}
\end{algorithm}\vspace{-0.2cm}
\subsection{Subproblem with respect to $\{\mathbf{U},\mathbf{F}\}$:}
In this section, we optimize $\mathbf{F}$ with fixed ${\mathbf{R}}_{s,u}$ and $\mathbf{w}_{k,u}$. Defining auxiliary matrix as follows: $\mathbf{Q}=\tilde{\mathbf{F}}\mathbf{\Phi}\mathbf{\Phi}^H\tilde{\mathbf{F}}^H$. The subproblem with respect to $\{\mathbf{U},\mathbf{F}\}$ is giving by\vspace{-0.1cm}
\begin{subequations}\label{problem:uplink_FQ}
    \begin{align}        
        \!\min_{\mathbf{U},\tilde{\mathbf{F}},\mathbf{Q}} \  &  \text{tr}(\mathbf{U}^{-1}) \\[-0.2cm]
        \label{constraint:PDD}
        \mathrm{s.t.} \  & \mathbf{Q} = \tilde{\mathbf{F}}\mathbf{\Phi}\mathbf{\Phi}^H\tilde{\mathbf{F}}^H, \\
        \label{constraint:FIM_PDD}
& \left[{{\bf{J}}_{11}(\mathbf{Q})-\mathbf{U}}\ \ {{\bf{J}}_{12}(\mathbf{Q})}; {{\bf{J}}_{12}^T(\mathbf{Q})}\ \ {{\bf{J}}_{22}(\mathbf{Q})}\right] \succeq 0, \\
        \label{constraint:analog_tilde}
        & \tilde{\mathbf{F}} \in \mathcal{A}_{\tilde{F}}, \\
        & \eqref{constraint:communication_SINR}, \eqref{constraint:U},
    \end{align}
\end{subequations}\vspace{-0.5cm}

\noindent where the matrices in the constraint \eqref{constraint:FIM_PDD} are obtained by substituting~\eqref{constraint:PDD} into results in Appendix~B. Specifically, they can be expressed as\vspace{-0.1cm}
\begin{equation}
\begin{aligned}
    \mathbf{J}_{11}& (\mathbf{Q}) =\frac{2 |\beta|^2 T }{\sigma_u^2}\times\\
    &\Re \left\{    \begin{bmatrix}
        \mathrm{tr}(\mathbf{Q} \dot{\mathbf{G}}_{\theta_{s}} \mathbf{R}_{u} \dot{\mathbf{G}}_{\theta_{s}}^H) &\mathrm{tr}(\mathbf{Q} \dot{\mathbf{G}}_{\theta_{s}} \mathbf{R}_{u} \dot{\mathbf{G}}_{r_s}^H )\\
        \mathrm{tr}(\mathbf{Q} \dot{\mathbf{G}}_{r_s} \mathbf{R}_{u} \dot{\mathbf{G}}_{\theta_{s}}^H )& \mathrm{tr}(\mathbf{Q} \dot{\mathbf{G}}_{r_s} \mathbf{R}_{u} \dot{\mathbf{G}}_{r_s}^H)
    \end{bmatrix}\right\},
\end{aligned}
\end{equation}
\begin{equation}
    \mathbf{J}_{12}(\mathbf{Q}) = \frac{2 T}{\sigma_u^2} \Re \left\{ \begin{bmatrix}
        \beta^* \mathrm{tr}( \mathbf{Q} \tilde{\mathbf{G}} \mathbf{R}_{u} \dot{\mathbf{G}}_{\theta_s}^H ) \\
        \beta^* \mathrm{tr}( \mathbf{Q}  \tilde{\mathbf{G}} \mathbf{R}_{u}  \dot{\mathbf{G}}_{r_s}^H )
    \end{bmatrix} [1, j] \right\},
\end{equation}
\begin{equation}
    \mathbf{J}_{22}(\mathbf{Q}) = \frac{2 T}{\sigma_u^2} \mathbf{I}_2 \mathrm{tr} ( \mathbf{Q} \tilde{\mathbf{G}} \mathbf{R}_{u} \tilde{\mathbf{G}}^H ).\vspace{-0.1cm}
\end{equation}
\noindent Similar to problem~\eqref{problem:downlink_U_PDD}, problem~\eqref{problem:uplink_FQ} is non-convex due to the non-convex SINR constraint~\eqref{constraint:communication_SINR}, quadratic constraint~\eqref{constraint:PDD}, and the unit modulus constraint~\eqref{constraint:analog_tilde}. We adopt the PDD optimization technique to solve this problem~\cite{9120361}. Specifically, by introducing the Lagrangian dual matrix ${\bm{\Upsilon}}$ and the penalty factor $\rho$ for constraint~\eqref{constraint:PDD}, the AL problem can be given as:\vspace{-0.1cm}
\begin{subequations}
    \begin{align}
        \label{problem:uplink_FQ_AL}
        \min_{\mathbf{U},\mathbf{Q},\tilde{\mathbf{F}}} \quad &  \text{tr}(\mathbf{U}^{-1})+\frac{1}{2\rho}\|\mathbf{Q}-\tilde{\mathbf{F}}\mathbf{\Phi}\mathbf{\Phi}^H\tilde{\mathbf{F}}^H-\rho{\bm{\Upsilon}} \|^2 \\[-0.1cm]
        \mathrm{s.t.} \quad & \eqref{constraint:communication_SINR}, \eqref{constraint:U},\eqref{constraint:FIM_PDD},\eqref{constraint:analog_tilde}.
    \end{align}
\end{subequations}\vspace{-0.5cm}

The proposed PDD-based algorithm solves problem \eqref{problem:uplink_FQ_AL} with a double-loop iteration. In the outer iteration loop, the Lagrangian dual matrix and penalty factor are updated. In the inner iteration loop, the AL problem is solved with the AO method. For the inner loop AO, we partition the optimization variables into two blocks, i.e., $\{\mathbf{U},{\mathbf{Q}}\}$ and ${\mathbf{F}}$. During each iteration, we successively optimize the variables within one block, while keeping the variables in the other block constant. 

\subsubsection{Subproblem with respect to $\{\mathbf{U},{\mathbf{Q}}\}$} The subproblem with respect to $\{\mathbf{U},{\mathbf{Q}}\}$ is given by\vspace{-0.1cm}
\begin{subequations}\label{problem:uplink_PDD_UQ}
    \begin{align}       
        \min_{\mathbf{U},\mathbf{Q}} \quad &  \text{tr}(\mathbf{U}^{-1})+\frac{1}{2\rho}\|\mathbf{Q}-\tilde{\mathbf{F}}\mathbf{\Phi}\mathbf{\Phi}^H\tilde{\mathbf{F}}^H-\rho{\bm{\Upsilon}} \|^2 \\[-0.1cm]
        \mathrm{s.t.} \quad & \eqref{constraint:U},\eqref{constraint:FIM_PDD}.
    \end{align}
\end{subequations}\vspace{-0.5cm}

\noindent Problem \eqref{problem:uplink_PDD_UQ} is convex and its optimal solution can be solved via CVX.
\subsubsection{Subproblem with respect to $\tilde{\mathbf{F}}$} The subproblem with respect to $\tilde{\mathbf{F}}$ is given by\vspace{-0.1cm}
\begin{subequations}\label{problem:uplink_PDD_F}
    \begin{align}        
        \min_{\tilde{\mathbf{F}}} \quad &  \|\mathbf{Q}-\tilde{\mathbf{F}}\mathbf{\Phi}\mathbf{\Phi}^H\tilde{\mathbf{F}}^H-\rho{\bm{\Upsilon}} \|^2 \\[-0.2cm]
        \mathrm{s.t.} \quad & \eqref{constraint:communication_SINR}, \eqref{constraint:analog_tilde}.
    \end{align}
\end{subequations}\vspace{-0.5cm}

\noindent The SINR constraint for user $k$ in \eqref{constraint:communication_SINR} can be rewritten as in~\eqref{constraint:communication_SINR_REFOR2} at the top of next page,
\begin{figure*}[!t]
\normalsize\vspace{-0.3cm}
\begin{equation}\label{constraint:communication_SINR_REFOR2}
\begin{aligned}
\frac{P_u}{\gamma_{k,u}} |\hat{\mathbf{w}}_{k,u}^H \tilde{\mathbf{F}}^H {\mathbf{h}}_{k}|^2  &- \hat{\mathbf{w}}_{k,u}^H \tilde{\mathbf{F}}^H {\mathbf{G}} \mathbf{R}_{u} {\mathbf{G}}^H \tilde{\mathbf{F}} \hat{\mathbf{w}}_{k,u} - \sum\nolimits_{i\neq k} P_u |\hat{\mathbf{w}}_{k,u}^H \tilde{\mathbf{F}}^H {\mathbf{h}}_{i}|^2 = \frac{P_u}{\gamma_k} |{\bm{f}}^H\hat{\mathbf{W}}_{k,u}^H  {\mathbf{h}}_{k}|^2  \\&- {\bm{f}}^H\hat{\mathbf{W}}_{k,u}^H {\mathbf{G}} \mathbf{R}_{u} {\mathbf{G}}^H \hat{\mathbf{W}}_{k,u} {\bm{f}}-  \sum\nolimits_{i\neq k} P_u |{\bm{f}}^H\hat{\mathbf{W}}_{k,u}^H {\mathbf{h}}_{i}|^2  =\text{tr} (\hat{\mathbf{F}}\mathbf{B}_{k})\le \sigma_u^2 \|\mathbf{w}_{k,u}\|^2.
\end{aligned}
\end{equation}
\hrulefill \vspace*{0pt}\vspace{-0.3cm}
\end{figure*}
where $\hat{\mathbf{w}}_{k,u}=\mathbf{\Phi}\mathbf{w}_{k,u} $,  $\hat{\mathbf{W}}_{k,u}=\text{diag}\left(\hat{\mathbf{w}}_{k,u}\right)$, $\hat{\mathbf{F}}={\bm{f}}{\bm{f}}^H$, and $\mathbf{B}_{k}=\frac{P_u}{\gamma_{k,u}}\hat{\mathbf{W}}_{k,u}^H  {\mathbf{h}}_{k}{\mathbf{h}}_{k}^H\hat{\mathbf{W}}_{k,u}-\sum_{i\neq k}P_u\hat{\mathbf{W}}_{k,u}^H {\mathbf{h}}_{i}{\mathbf{h}}_{i}^H\hat{\mathbf{W}}_{k,u}-\hat{\mathbf{W}}_{k,u}^H {\mathbf{G}} \mathbf{R}_{u} {\mathbf{G}}^H \hat{\mathbf{W}}_{k,u}$. Substituting \eqref{Phi} into the objective function of problem~\eqref{problem:uplink_PDD_F}, we have \vspace{-0.1cm}
\begin{equation}
\tilde{\mathbf{F}}\mathbf{\Phi}\mathbf{\Phi}^H\tilde{\mathbf{F}}^H=\tilde{\mathbf{F}}\sum\nolimits_{i=1}^{N_{RF}}\bm{\theta}_i\bm{\theta}_i^H\tilde{\mathbf{F}}^H=\sum\nolimits_{i=1}^{N_{RF}}\mathbf{\Phi}_i\hat{\mathbf{F}}\mathbf{\Phi}_i^H,\vspace{-0.1cm}
\end{equation}
where $\mathbf{\Phi}_i=\text{diag}\left(\bm{\theta}_i\right)$. Note that the definition of $\hat{\mathbf{F}}$ implies $\hat{\mathbf{F}} \succeq 0$ and $\text{Rank}(\hat{\mathbf{F}})=1$. Then, problem \eqref{problem:uplink_PDD_F} can be reformulated into following form\vspace{-0.1cm}
\begin{subequations}\label{problem:uplink_PDD_F_SDR}
    \begin{align}       
        \min_{\hat{\mathbf{F}}} \quad &  \|\mathbf{Q}-\sum\nolimits_{i=1}^{N_{RF}}\mathbf{\Phi}_i\hat{\mathbf{F}}\mathbf{\Phi}_i^H-\rho{\bm{\Upsilon}} \|^2 \\[-0.2cm]
        \label{constraint:SINR_F}
        \mathrm{s.t.} \quad & \text{tr} (\hat{\mathbf{F}}\mathbf{B}_{k})-\sigma_u^2 \|\mathbf{w}_{k,u}\|^2\le 0, \forall k \in \mathcal{K}, \\
        \label{constraint:modu_F}
        & [\hat{\mathbf{F}}]_{n,n} =1 , \forall n = 1, 2, \cdots, N_a, \\
        \label{constraint:SEMIdefinit_F}
        & \hat{\mathbf{F}} \succeq 0, \text{rank}(\hat{\mathbf{F}}) =1.
    \end{align}
\end{subequations}\vspace{-0.5cm}

\noindent The SDR problem of problem \eqref{problem:uplink_PDD_F_SDR} is convex and its optimal solution can be solved via CVX. Despite the SDR could lead to solutions with general rank, the Gaussian randomization can be used to further obtain rank-one solutions. The proposed algorithm for solving the subproblem with respect to $\mathbf{F}$ is summarized in \textbf{Algorithm~\ref{algorithm2}}. \vspace{-0.2cm}
\begin{algorithm}[htbp]
\caption{PDD-based algorithm for solving problem~\eqref{problem:uplink_FQ}}\label{algorithm2}
\begin{algorithmic}[1]
\STATE {Initialize feasible ${\bm{\Upsilon}}^{(n)}$, $\eta^{(n)}$, and ${\rho}^{(n)}$ with $n=0$, set $\mu \in (0,1)$.}
\STATE {\bf repeat: outer loop}
\STATE \quad Obtain AL problem ~\eqref{problem:uplink_FQ_AL} with respect to ${\bm{\Upsilon}}^{(n)}$ and~${\rho}^{(n)}$.\STATE \quad {\bf repeat: inner loop}
\STATE \quad\quad Given $\mathbf{F}=\tilde{\mathbf{F}}\mathbf{\Phi}$, update $\mathbf{Q}$ by solving problem~\eqref{problem:uplink_PDD_UQ}.
\STATE \quad\quad Given $\mathbf{Q}$, update $\mathbf{F}=\tilde{\mathbf{F}}\mathbf{\Phi}$ by solving problem~\eqref{problem:uplink_PDD_F}.
\STATE \quad {\bf until} the fractional decrease of~\eqref{problem:uplink_FQ_AL} is less than threshold $\epsilon_{\ref{algorithm2},\text{AO}}$. 
\STATE \quad Obtain $\mathbf{F}^{(n+1)}=\tilde{\mathbf{F}}^{(n+1)}\mathbf{\Phi}$ and $\mathbf{Q}^{(n+1)}$.
\STATE \quad {\bf if} $\|\mathbf{Q}^{(n+1)}-\tilde{\mathbf{F}}^{(n+1)}\mathbf{\Phi}\mathbf{\Phi}^H(\tilde{\mathbf{F}}^{(n+1)})^H\|_\infty \le \eta^{(n)}$ {\bf then}
\STATE \quad \quad Update ${\bm{\Upsilon}}$ with ${\bm{\Upsilon}}^{(n+1)}={\bm{\Upsilon}}^{(n)}+\frac{1}{\rho^{(n)}}(\mathbf{Q}^{(n+1)}-\tilde{\mathbf{F}}^{(n+1)}\mathbf{\Phi}\mathbf{\Phi}^H(\tilde{\mathbf{F}}^{(n+1)})^H)$.
\STATE \quad \quad Penalty factor remains the same, i.e., $\rho^{(n+1)}=\rho^{(n)}$.\STATE \quad {\bf else}
\STATE \quad \quad Update $\rho$ with $\rho^{(n+1)}=\mu\rho^{(n)}$.
\STATE \quad \quad Lagrangian dual matrix ${\bm{\Upsilon}}$ remains the same, i.e., ${\bm{\Upsilon}}^{(n+1)}={\bm{\Upsilon}}^{(n)}$.
\STATE \quad {\bf end if}
\STATE \quad Set $\eta^{(n+1)}=0.9\|\mathbf{Q}^{(n+1)}-\tilde{\mathbf{F}}^{(n+1)}\mathbf{\Phi}\mathbf{\Phi}^H(\tilde{\mathbf{F}}^{(n+1)})^H\|_\infty$, $n=n+1$.
\STATE {\bf until} the constraint violation is less than threshold $\epsilon_{\ref{algorithm2},\text{PDD}}$.
\end{algorithmic}
\end{algorithm} \vspace{-0.2cm}
\subsection{The Overall Algorithm for Solving~\eqref{problem:uplink_U}}
The proposed algorithm for solving~\eqref{problem:uplink_U} during the uplink ISPAC is summarized in \textbf{Algorithm~\ref{algorithm3}}. Every iteration of \textbf{Algorithm~\ref{algorithm3}} yields a non-increasing objective function value for~\eqref{problem:uplink_U}. Furthermore, because the BS power budget is constrained, the objective function value of~\eqref{problem:uplink_U} is lower bounded. Therefore, it is guaranteed that \textbf{Algorithm~\ref{algorithm3}} will converge to a stationary point of~\eqref{problem:uplink_U}. The main complexity of \textbf{Algorithm~\ref{algorithm3}} comes from solving~\eqref{problem:uplink_RW} and~\eqref{problem:uplink_FQ} with \textbf{Algorithm~\ref{algorithm1}} and \textbf{Algorithm~\ref{algorithm2}}. Given the solution accuracy $\epsilon_{\ref{algorithm1},\text{SCA}}$, $\epsilon_{\ref{algorithm2},\text{PDD}}$, and $\epsilon_{\ref{algorithm3},\text{AO}}$, the complexity of solving problem~\eqref{problem:uplink_U} with the interior-point method is $\mathcal{O}(\log(1/\epsilon_{\ref{algorithm3},\text{AO}})(\log(1/\epsilon_{\ref{algorithm1},\text{SCA}})N_{RF}^{4.5}\!+\!\log(1/\epsilon_{\ref{algorithm2},\text{PDD}})N_a^{4.5}))$~\cite{5447068}.
\begin{algorithm}[htbp]
\caption{AO algorithm for solving problem~\eqref{problem:uplink_U}}\label{algorithm3}
\begin{algorithmic}[1]
\STATE {Initialize the analog matrix $\mathbf{F}$ with random phases.}
\STATE {{\bf{repeat:}}}
\STATE {\quad Given $\mathbf{F}$, update $\mathbf{R}_{u}$ and $\mathbf{w}_{k,u}, \forall k\in \mathcal{K}$ by solving problem~\eqref{problem:uplink_RW} with algorithm~\ref{algorithm1}.}
\STATE {\quad Given $\mathbf{R}_{u}$ and $\mathbf{w}_{k,u}, \forall k\in \mathcal{K}$, update $\mathbf{F}$ by solving~\eqref{problem:uplink_FQ} with algorithm~\ref{algorithm2}.}
\STATE  {\bf until} the fractional decrease of~\eqref{problem:uplink_U} is less than the threshold $\epsilon_{\ref{algorithm3},\text{AO}}$.
\end{algorithmic}
\end{algorithm}

\section{Numerical Results} \label{sec:results}

In this section, we provide the numerical results obtained by Monte Carlo simulations to verify the effectiveness of the proposed algorithms. \vspace{-0.2cm}
\subsection{Simulation Setup}
It is assumed that the ISPAC system serves $K=4$ communication users while simultaneously carrying out target sensing and positioning at a frequency of $28$ GHz. The MT and AT are both equipped with ULA antenna arrays. The array aperture of the MT is $0.5$ m, which leads to a Rayleigh distance of around $50$ m. All communication users and the target are located in the near-field region of the MT. Specifically, the target is set at the direction of $45^\circ$ with a distance of $20$ m. Communication users as well as scatterers for communication links fall in the distance from $20$ m to $30$ m with respect to the MT at random angles following the uniform distribution. The number of scatterers associated with each user is set to be $L_k=2, \forall k\in\mathcal{K}$. Path loss coefficients are calculated utilizing the Empirical NYC path loss model~\cite{6834753}, which is given by:\vspace{-0.1cm}
\begin{equation}
\begin{aligned}
    L(r)[{\text{dB}}] = a_1 + a_210\log_{10}(r),
\end{aligned}\vspace{-0.1cm}
\end{equation}
where $r$ is the propagation distance. $a_1$ and $a_2$ are the path loss at the reference distance and the path loss exponent, respectively. Based on the measurement at the frequency of $28$ GHz, parameters $\{a_1,a_2\}$ are set as $\{61.4,2\}$ for LoS channels and $\{72,2.92\}$ for NLoS channels~\cite{6834753}. The power budget at the BS and users are set as $30$ dBm and $20$ dBm, respectively. The noise power at the BS and users are set as $-80$ dBm. The reflection coefficient of the target, i.e., ${\beta}_r$, is set to $0$ dB with random phase. This means the target is assumed to reflect all probing signals impinging on it. The QoS requirements of all communication users are set as the same value, i.e., $\gamma_{k,d}=\gamma_d, \gamma_{k,u}=\gamma_u, \forall k\in \mathcal{K}$. All simulation results in this section are obtained with an average of 100 channel realizations unless otherwise specified. In simulation figures, legends ``HB'', ``FD'', ``near'', and ``far'' represent fully digital structure at the MT, hybrid analog and digital structure at the MT, near-field beamfocusing for communication, and far-field beamsteering for communication, respectively. ``RCRB'' and ``RMSE'' represent the root of CRB and MSE, respectively.\vspace{-0.2cm}
\subsection{Baseline Schemes}
We compare with the following two baseline schemes to verify the efficacy of the proposed ISPAC framework in both downlink and uplink.
\begin{enumerate}
\item \textbf{Far-field beamsteering:} In this baseline scheme, the far-field channel model is adopted for communication channels since the near-field channel state information (CSI) is more difficult to obtain than the far-field CSI. The channel between the MT and user $k$ is
\vspace{-0.1cm}
\begin{equation}\label{channel_h_k_far}
{\mathbf{h}}_{k,\text{far}}=\alpha_k\mathbf{e}^{N_a}_\text{far}\left(\theta_k\right)+\sum\nolimits_{l=1}^{L_k}\frac{1}{\sqrt{L_k}}\alpha_l^k\mathbf{e}^{N_a}_\text{far}\left(\theta_l^k\right),\vspace{-0.1cm}
\end{equation} 
where the channel is represented with far-field array response vectors,  i.e., $\mathbf{e}^{N_a}_\text{far}\left(\cdot\right)$. As the near-field communication channel in~\eqref{channel_h_k}, the far-field communication channel also includes the LoS part and NLoS part introduced by $L_k$ scatterers. Moreover, the coefficients $\alpha_k$, $\alpha_l^k$, $\forall k=1,2,\cdots,K$, $\forall l=1,2,\cdots,L_k$ are set identical to its near-field counterpart in \eqref{channel_h_k} for fair comparison.
\item \textbf{Fully digital ISPAC (CRB lower bound):} In this baseline scheme, the MT adopts FD precoding structure where the number of RF chains $N_{RF}$ is equal to the number of antennas $N_a$. This scheme provides the CRB lower bound for our proposed ISPAC framework. 
\end{enumerate}\vspace{-0.4cm}
\subsection{Convergence Behavior of Proposed Algorithms}
\begin{figure}[!htbp]
\vspace{-0.5cm}
\centering
\begin{minipage}[t]{0.49\textwidth}
\centering
\includegraphics[width=2.5in]{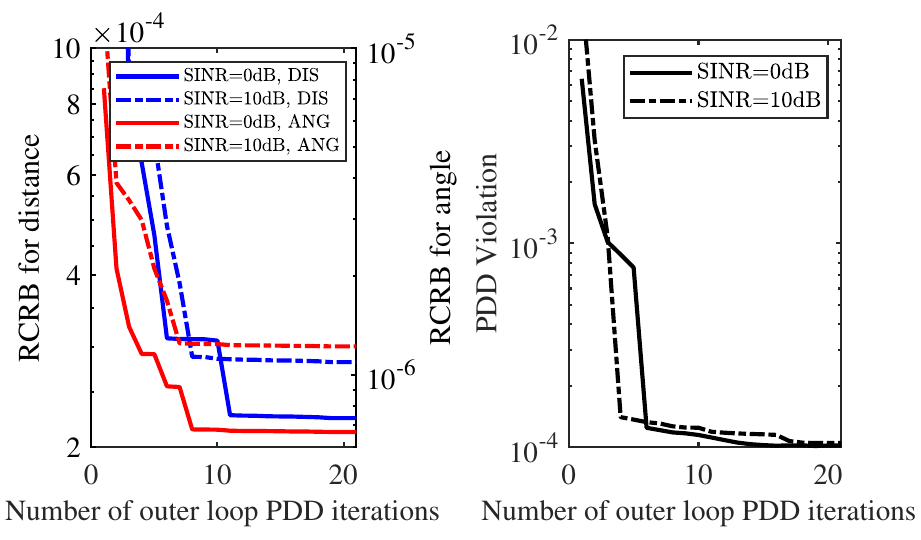}\vspace{-0.2cm}
\caption{Convergence behavior of \bf{Algorithm~\ref{algorithm4}}.}\label{fig:convergence_downlink}
\end{minipage}
\begin{minipage}[t]{0.49\textwidth}
\centering
\includegraphics[width=2.5in]{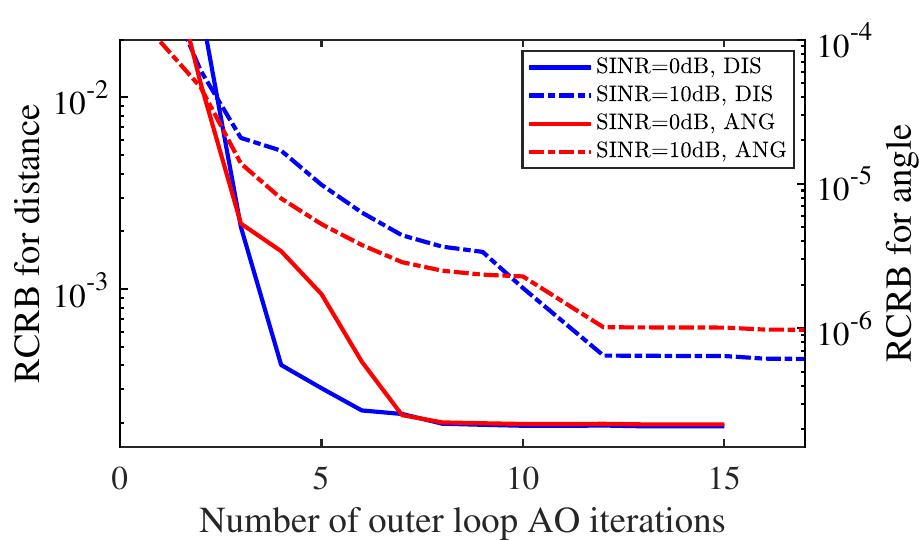}\vspace{-0.2cm}
\caption{Convergence behavior of \bf{Algorithm~\ref{algorithm3}}.}\label{fig:convergence_uplink}\vspace{-0.3cm}
\end{minipage}
\end{figure}
In Fig.~\ref{fig:convergence_downlink} and~\ref{fig:convergence_uplink}, we examine the convergence performance of proposed algorithms for the downlink and uplink ISPAC, respectively. Parameters $\{N_a, N_b\}$ are set as $\{64, 16\}$ and the sub-array size of the HAD structure is $M = 4$. Results in this subsection are obtained from a random channel realization. 

For downlink ISPAC, the convergence behaviour of the proposed PDD-based \textbf{Algorithm~\ref{algorithm4}} is shown in Fig.~\ref{fig:convergence_downlink}. The convergence thresholds of the inner iteration and the outer iteration in \textbf{Algorithm~\ref{algorithm4}} are set as $\epsilon_{\ref{algorithm4},\text{AO}}=10^{-3}$ and $\epsilon_{\ref{algorithm4},\text{PDD}}=10^{-4}$. The PDD constraint violation for \textbf{Algorithm~\ref{algorithm4}} is shown in the right of Fig.~\ref{fig:convergence_downlink}.  As can be seen, as the number of outer loop iterations rises, the constraint violation rapidly drops until it meets the predetermined accuracy. This indicates feasible $\mathbf{F}$, $\mathbf{R}_{s,d}$, and $\mathbf{W}$ are obtained with \textbf{Algorithm~\ref{algorithm4}}.

In terms of uplink ISPAC, the convergence behaviour of the proposed AO \textbf{Algorithm~\ref{algorithm3}} is given in Fig.~\ref{fig:convergence_uplink}. Specifically, in simulations, the convergence thresholds of the SCA method, PDD-based algorithm, and AO algorithm involved in \textbf{Algorithm~\ref{algorithm1}}, \textbf{Algorithm~\ref{algorithm2}}, and \textbf{Algorithm~\ref{algorithm3}}  are given as $\epsilon_{\ref{algorithm1},\text{SCA}}=\epsilon_{\ref{algorithm2},\text{AO}}=\epsilon_{\ref{algorithm3},\text{AO}}=10^{-3}$ and $\epsilon_{\ref{algorithm2},\text{PDD}}=10^{-4}$. It can be seen that the proposed algorithm can converge within $20$ iterations when the ISPAC system works under the uplink working mode.\vspace{-0.2cm}
\subsection{Target Location Estimation}
\begin{figure}[ht]\vspace{-0.3cm}
\centering
\subfigure[Downlink angle and distance estimation.]{\label{fig:MUSIC_down}
\includegraphics[width=2.5in]{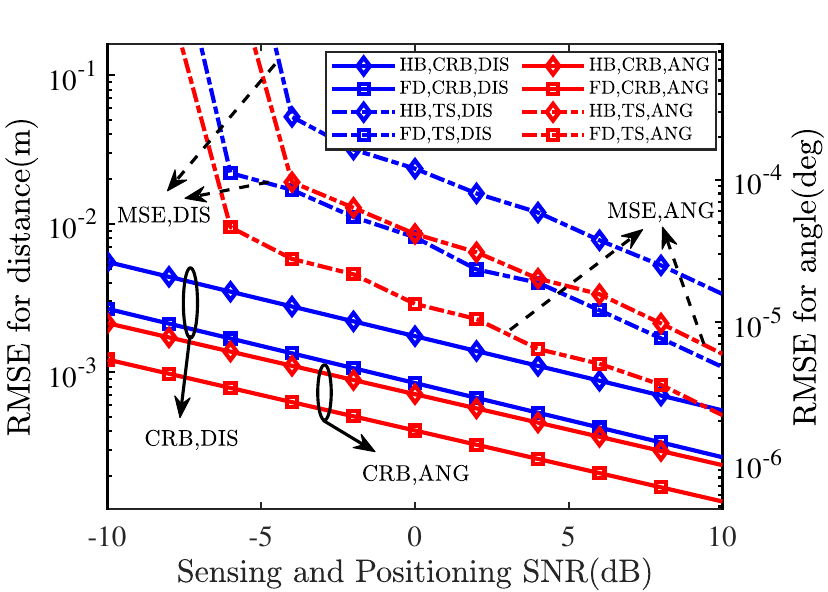}}\vspace{-0.2cm}
\subfigure[Uplink angle and distance estimation.]{\label{fig:MUSIC_up}
\includegraphics[width=2.5in]{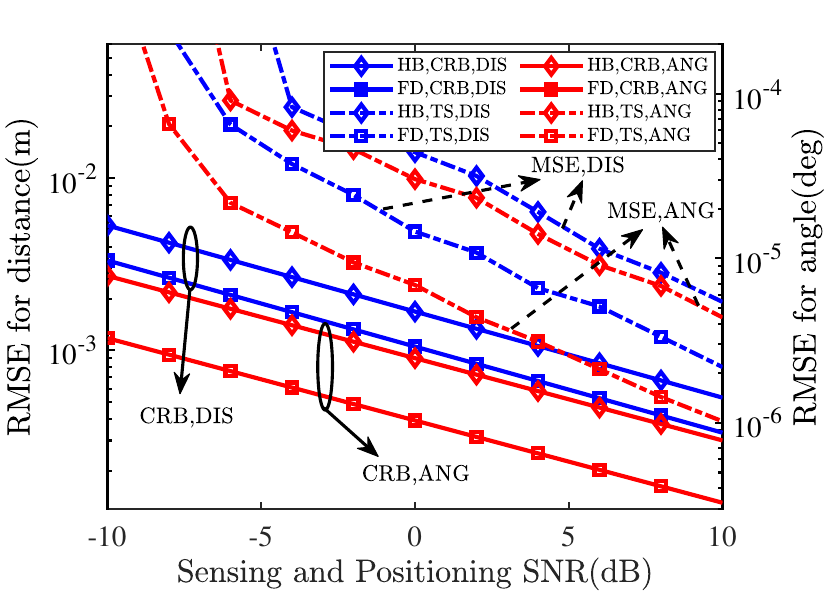}}\\\vspace{-0.2cm}
\caption{Target positioning performance versus the sensing SNR.}\label{fig:MUSIC}\vspace{-0.2cm}
\end{figure}
In Fig.~\ref{fig:MUSIC}, we show the target positioning performance versus the sensing SNR in terms of the root-MSE (RMSE) of target location estimation. Specifically, parameters $\{N_a, N_b\}$ is set as $\{64, 16\}$. The sub-array size of the HAD structure is $M = 4$. The communication QoS requirement is set as $10$ dB.  The target location is estimated using the proposed uplink positioning algorithm and downlink positioning algorithm. The sensing SNR is defined as ${|\beta_s|^2P_d}/{\sigma_d^2}$ and ${|\beta_s|^2P_s}/{\sigma_u^2}$ for downlink and uplink ISPAC, respectively. As expected, the RMSE for target positioning is lower-bounded by the corresponding CRB. It can be verified that the proposed low complexity positioning algorithms can locate the target simultaneously in angle and distance domains for both for downlink and uplink ISPAC.\vspace{-0.2cm}
\subsection{Root RCB of Downlink ISPAC}
\begin{figure*}[ht]\vspace{-0.8cm}
\centering\hspace{-10mm}
\subfigure[RCRB of distance versus QoS.]{\label{fig:RCRB_DIS_SINR}
\includegraphics[width=2.5in]{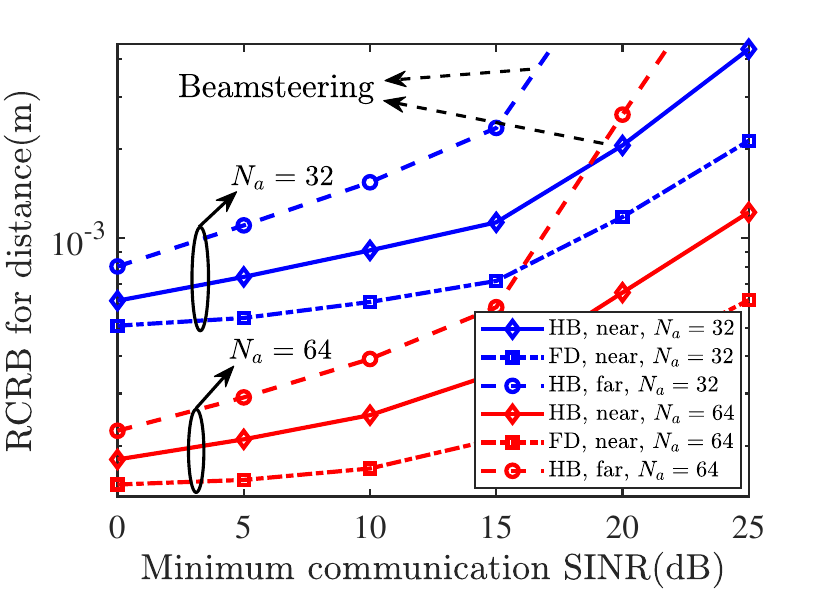}}\hspace{-5mm}
\subfigure[RCRB of angle versus QoS.]{\label{fig:RCRB_ANG_SINR}
\includegraphics[width=2.5in]{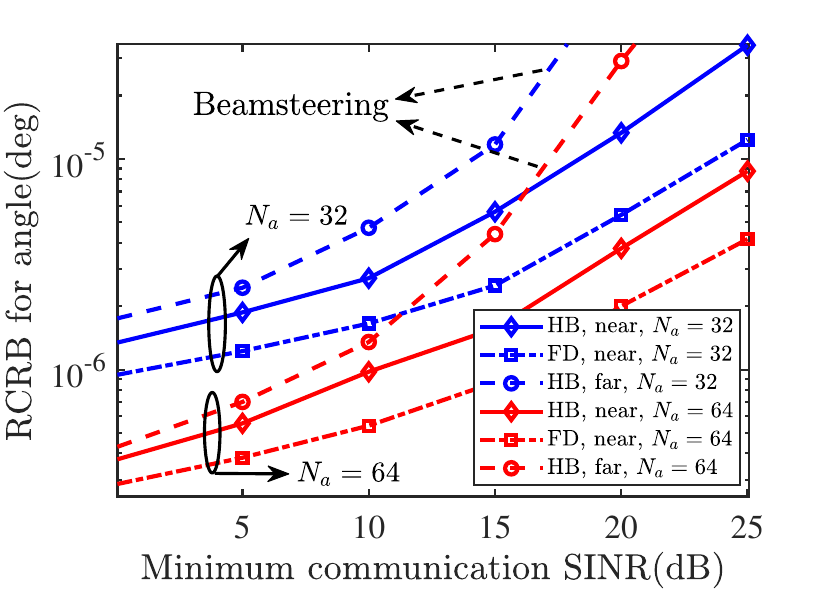}}\hspace{-5mm}
\subfigure[Downlink RCRB of target versus RF number.]{\label{fig:RCRB_RF}
\includegraphics[width=2.5in]{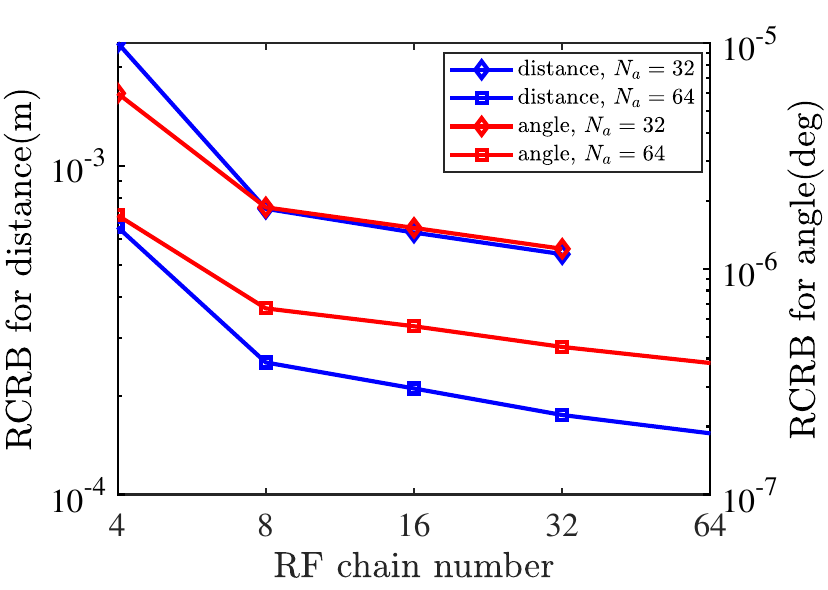}}\vspace{-0.2cm}\hspace{-5mm}
\caption{The RCRB of target positioning versus the QoS of users during downlink ISPAC.}\label{fig:RCRB_DOWN}\vspace{-0.2cm}
\end{figure*}
\begin{figure*}[ht]
\centering\hspace{-10mm}
\subfigure[RCRB of distance versus QoS.]{\label{fig:RCRB_DIS_SINR_UP}
\includegraphics[width=2.5in]{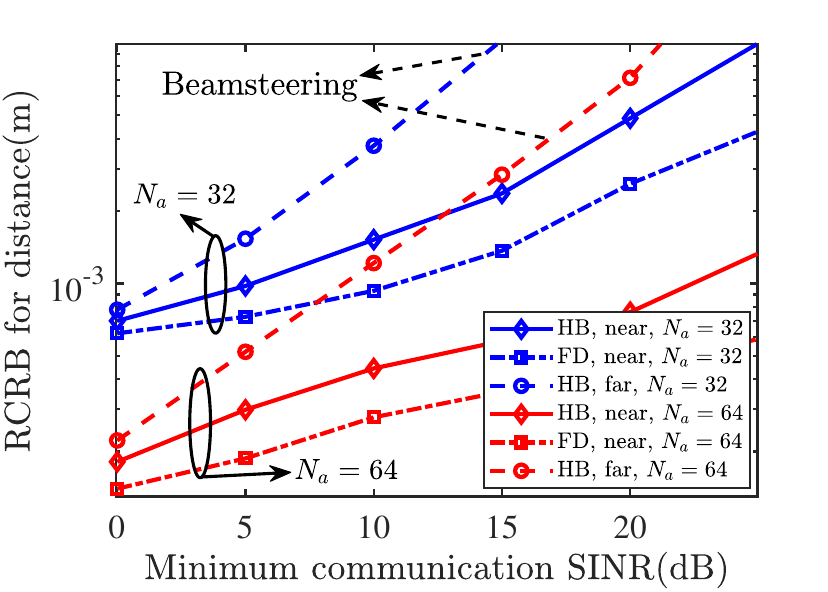}}\hspace{-5mm}
\subfigure[RCRB of angle versus QoS.]{\label{fig:RCRB_ANG_SINR_UP}
\includegraphics[width=2.5in]{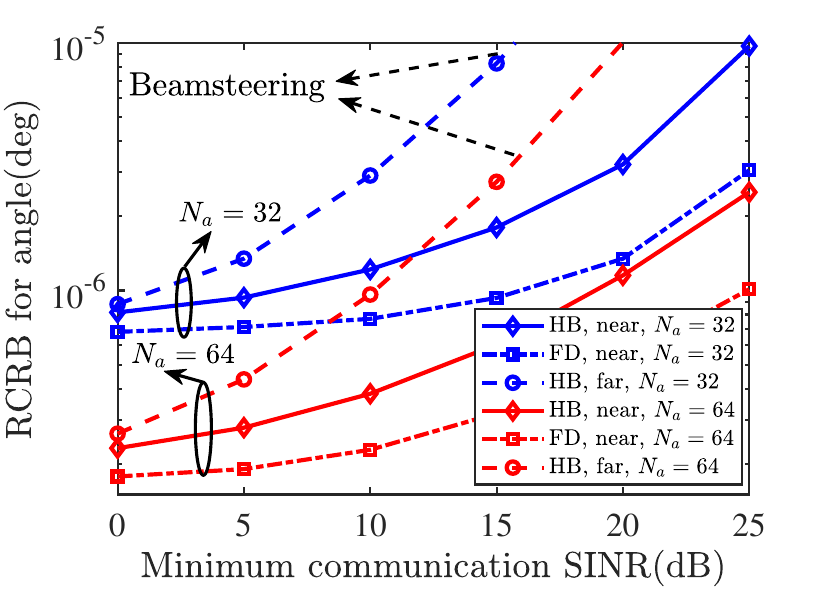}}\hspace{-5mm }
\subfigure[RCRB of target versus RF number.]{\label{fig:RCRB_RF_UP}
\includegraphics[width=2.5in]{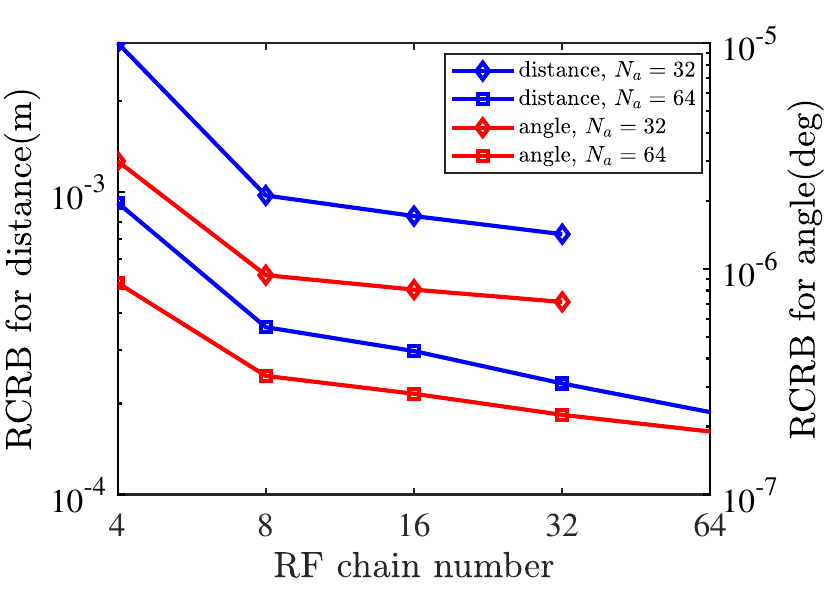}}\vspace{-0.2cm}\hspace{-5mm}
\caption{The RCRB of target positioning versus the QoS of users during uplink ISPAC.}\label{fig:RCRB_UP}\vspace{-0.2cm}
\end{figure*}
In Fig.~\ref{fig:RCRB_DOWN}, we investigate the RCRB for target positioning in terms of the QoS of users and the number of RF chains at the MT to demonstrate the trade-off between positioning and communication. Fig.~\ref{fig:RCRB_DIS_SINR} and~\ref{fig:RCRB_ANG_SINR} sketch the RCRB for target positioning versus the QoS of users. Two antenna configurations for MT and AT are considered in the simulation. Specifically, parameters $\{N_a, N_b\}$ are set as $\{32, 8\}$ or $\{64, 16\}$. The sub-array size of the HAD structure is set as $M=4$. Our observations indicate that as the QoS of communication users improves, the precision of target positioning diminishes. This phenomenon matches with the analysis in {\textbf{Remark~\ref{remark2}}}, i.e., as the QoS demand becomes more stringent, the design of HAD precoding matrix has a tendency to meet QoS demand rather than to achieve accurate target positioning.

With the same antenna configuration, the near-field ISPAC scheme always outperforms baseline \textbf{Scheme 1}, where the communication design is based on far-field beamsteering instead of near-field beamfocusing. The reason behind this is that the beamfocusing based on the near-field channel model can cast transmitted signals to both the intended angle and distance, i.e., a specific area. In contrast, the beamsteering based on the far-field channel model only supports sending signals toward the intended direction, i.e., a specific angle. Consequently, sophisticated beamfocusing leads to less inter-user interference among communication users compared to coarse beamsteering, which alleviates the stress of meeting user QoS demand and allows more DoFs for target positioning. \textbf{Scheme 2} with FD beamfocusing structure MT serves as a theoretical lower bound for the proposed hybrid beamfocusing structure since it introduces fewer constraints to system design. 

Fig.~\ref{fig:RCRB_RF} shows the RCRB for target positioning versus the number of RF chains at the MT. Parameters $\{N_a, N_b\}$ are set as $\{32, 8\}$ or $\{64, 16\}$. The users' QoS demand is set as $10$ dB. It can be observed that the RCRBs for both target angle and distance decrease when the MT is equipped with an increasing number of RF chains. This can be explained from two perspectives. In the downlink ISPAC system, the MT simultaneously bears the tasks of transmitting communication signals and probing signals. On the one hand, an increased number of RF chains provides more DoFs for the downlink beamfocusing and thus mitigates the inter-user interference among communication users; On the other hand, with an increased number of RF chains, more sophisticated probing signal design can be achieved to realize better target positioning performance. This observation is also in line with the results in Fig.~\ref{fig:RCRB_DIS_SINR} and~\ref{fig:RCRB_ANG_SINR}, where the \textbf{Scheme 2} gives the RCRB lower bound of the proposed scheme.\vspace{-0.2cm}
\subsection{Root RCB of Uplink ISPAC}
In Fig.~\ref{fig:RCRB_UP}, we investigate the RCRB for target positioning in terms of the QoS of users and the number of RF chains at the MT to demonstrate the trade-off between positioning and communication. Fig.~\ref{fig:RCRB_DIS_SINR_UP} and~\ref{fig:RCRB_ANG_SINR_UP} sketch the RCRB for target positioning versus the QoS of users. Antenna configurations for both MT and AT are identical to those in Fig.~\ref{fig:RCRB_DIS_SINR_UP} and~\ref{fig:RCRB_ANG_SINR_UP}. Our observations indicate that as the QoS of communication users improves, the precision of target positioning diminishes. This phenomenon can be attributed to the rigorous QoS demands, which necessitate the design of analog combiner with a tendency to meet QoS requirements. This observation is consistent with the UPC trade-off analyzed in {\textbf{Remark~\ref{remark1}}}.

The near-field ISPAC method outperforms baseline \textbf{Scheme 1}, where the communication design is based on far-field receiving beamsteering rather than near-field receiving beamfocusing. This is because the parallel-wave-based far-field channel solely uses angle information, leading to a high degree of correlation between the channels of users with similar angles. The BS is unable to regulate IUI effectively. In contrast, the spherical-wave-based near-field channel carries both the angle and distance information. The additional distance information aids in interference reduction at the BS side. \textbf{Scheme 2} with the FD MT provides a theoretical CRB lower bound for the proposed ISPAC framework since it places fewer restrictions on system design.

Fig.~\ref{fig:RCRB_RF_UP} shows the RCRB for target positioning versus the number of RF chains at the MT. Antenna configurations for both MT and AT are identical to those in Fig.~\ref{fig:RCRB_RF_UP}. The users' QoS demand is set as $10$ dB. It can be seen, for both antenna configurations, as the MT gets equipped with more RF chains, the RCRB for both target angle and distance decreases. This is to be expected as more RF chains provide the analog combining matrix design more DoFs and a more flexible UPC trade-off can be achieved. This observation is consistent with \textbf{Remark~\ref{remark1}}. \vspace{-0.1cm}

\section{Conclusions} \label{sec:conclusion}

A novel near-field ISPAC framework was proposed, where a double-array structure BS supports communication users and detects a target at the same time. Low-complexity positioning algorithms were conceived for target positioning. Effective joint angle and distance CRB optimization frameworks and target positioning algorithms were proposed for both downlink and uplink ISPAC. Numerical results confirmed that our proposed ISPAC system can estimate not only the angle but also the distance of the target. Besides, the HAD structure at the MT has little impact on positioning performance given the communication QoS demand is not strict. Furthermore, adapting near-field beamfocusing could enhance the sensing and positioning performance of ISPAC. 

\section*{Appendix~A\\Derivation of the FIM for Downlink ISPAC} \label{appendiex:FIM_dowm}
During downlink ISPAC, the probing signal collected by the AT, i.e., $\mathbf{y}_{s,d}=\text{vec}(\mathbf{Y}_{s,d})$ follows the Gaussian distribution $ \mathcal{CN}(\mathbf{u}, \mathbf{R}_n)$, where $\mathbf{u}=\text{vec}(\mathbf{G}^T\mathbf{X})$ and $\mathbf{R}_n = \sigma_d^2 \mathbf{I}_{N_bT}$. The element at the $\ell$-th row and the $p$-th column of $\mathbf{J}_{\boldsymbol{\eta}}$ can be calculated by~\cite{kay1993fundamentals}\vspace{-0.1cm}
\begin{equation} \label{eqn:FIM_point}
    [\mathbf{J}_{\boldsymbol{\eta}}]_{\ell,p} = \frac{2}{\sigma^2} \Re \left\{ \frac{\partial \mathbf{u}^H}{\partial \eta_\ell} \frac{\partial \mathbf{u}}{\partial \eta_p} \right\},\vspace{-0.1cm}
\end{equation} 
where $\eta_\ell$ denotes the $\ell$-th element of $\boldsymbol{\eta}$. With $\tilde{\mathbf{G}} = \mathbf{a}(\theta_s,r_s) \mathbf{b}^T(\theta_s)$, we have
\begin{align}
    &\frac{\partial \mathbf{u}}{\partial \theta_s} = \beta_s \mathrm{vec}(\dot{\mathbf{G}}_{\theta_s}^T \mathbf{X}),\  \frac{\partial \mathbf{u}}{\partial r_s} = \beta_s \mathrm{vec}(\dot{\mathbf{G}}_{r_s}^T \mathbf{X}),\\
    &\frac{\partial \mathbf{u}}{\partial \beta_s^r} = \mathrm{vec}(\tilde{\mathbf{G}}^T \mathbf{X} ),\  \frac{\partial \mathbf{u}}{\partial \beta_s^i} = j \cdot \mathrm{vec}(\tilde{\mathbf{G}}^T \mathbf{X} ),
\end{align}
where $\dot{\mathbf{G}}_{\theta_s}  = \frac{\partial \tilde{\mathbf{G}}}{\partial \theta_s} =  \frac{\partial \mathbf{a}}{{\partial \theta_s}} \mathbf{b}^T
    +  \mathbf{a} \frac{\partial \mathbf{b}^T}{\partial \theta_s}, \dot{\mathbf{G}}_{r_s} = \frac{\partial \tilde{\mathbf{G}}}{\partial r_s} =  \frac{\partial \mathbf{a}}{{\partial r_s}} \mathbf{b}^T$. 
For simplicity, $\theta_{s}$ and $r_{s}$ are dropped In the above formulas. Denoting the matrix $\mathbf{J}_{11}$ as $\left[J_{\theta_{s} \theta_{s}} \ \  J_{\theta_{s} r_s}; J_{r_s \theta_{s}} \ \  J_{r_s r_s}\right]$, its entries can be given as follows:\vspace{-0.1cm}
\begin{equation} \label{eqn:FIM_element_down}
    J_{l p} =  \frac{2 |\beta|^2 T }{\sigma_d^2}   \Re \{ \mathrm{tr}(  \dot{\mathbf{G}}_{p}^T \mathbf{F}\tilde{\mathbf{R}}_x\mathbf{F}^H \dot{\mathbf{G}}_{l}^*  ) \}.\vspace{-0.1cm}
\end{equation}
where $\tilde{\mathbf{R}}_x=\mathbf{W}\mathbf{W}^H+\mathbf{R}_s$.
Next, the matrices $\mathbf{J}_{12}$ and $\mathbf{J}_{22}$
are derived as follows:
\begin{align}\vspace{-0.1cm}
    \mathbf{J}_{12} &= \frac{2 T}{\sigma_d^2} \Re \left( \begin{bmatrix}
        \beta^* \mathrm{tr}(  \tilde{\mathbf{G}}^T \mathbf{F}\tilde{\mathbf{R}}_x\mathbf{F}^H \dot{\mathbf{G}}_{\theta_s}^* )\\
        \beta^* \mathrm{tr}(  \tilde{\mathbf{G}}^T \mathbf{F}\tilde{\mathbf{R}}_x\mathbf{F}^H  \dot{\mathbf{G}}_{r_s}^* )
    \end{bmatrix} [1, j] \right),
    \end{align}\vspace{-0.1cm}
    \begin{align}
    \mathbf{J}_{22} = \frac{2 T}{\sigma_d^2} \mathbf{I}_2 \mathrm{tr} (  \tilde{\mathbf{G}}^T \mathbf{F}\tilde{\mathbf{R}}_x\mathbf{F}^H \tilde{\mathbf{G}}^*  ).\vspace{-0.3cm}
\end{align}

\section*{Appendix~B\\Derivation of the FIM for Uplink ISPAC} \label{appendiex:FIM}
During uplink ISPAC, the probing signal collected by the MT over $T$ time slots, i.e., $\mathbf{y}_{s,u}=\text{vec}(\mathbf{Y}_{s,u})$, follows the Gaussian distribution $\mathcal{CN}(\mathbf{u}, \mathbf{R}_n)$, where $\mathbf{u}=\text{vec}(\mathbf{F}^H\mathbf{G}\mathbf{S}_u)$ and $\mathbf{R}_n = \sigma_u^2 \mathbf{I}_{N_{RF}T}$. Similar to the derivation in Appendix~A, the entries $J_{l p}, \forall l, p \in \{\theta_{s}, r_s\}, $ of the matrix $\mathbf{J}_{11}$ can be given as follows:
\begin{align} \label{eqn:FIM_element}
    \!\!J_{l p} & = \frac{2 |\beta|^2 T }{\sigma_u^2}   \Re \{ \mathrm{tr}( \mathbf{F}^H \dot{\mathbf{G}}_{p} \mathbf{R}_{u} \dot{\mathbf{G}}_{l}^H \mathbf{F} ) \}, \forall l, p \in \{\theta_{s}, r_s\}.
\end{align}
Next, the matrices $\mathbf{J}_{12}$ and $\mathbf{J}_{22}$
are derived as follows:
\begin{align}
    \mathbf{J}_{12} &= \frac{2 T}{\sigma_u^2} \Re \left\{ \begin{bmatrix}
        \beta^* \mathrm{tr}( \mathbf{F}^H \tilde{\mathbf{G}} \mathbf{R}_{u} \dot{\mathbf{G}}_{\theta_s}^H \mathbf{F}) \\
        \beta^* \mathrm{tr}( \mathbf{F}^H  \tilde{\mathbf{G}} \mathbf{R}_{u}  \dot{\mathbf{G}}_{r_s}^H \mathbf{F})
    \end{bmatrix} [1, j] \right\},
    \end{align}\vspace{-0.1cm}
    \begin{align}
    \mathbf{J}_{22} = \frac{2 T}{\sigma_u^2} \mathbf{I}_2 \mathrm{tr} ( \mathbf{F}^H \tilde{\mathbf{G}} \mathbf{R}_{u} \tilde{\mathbf{G}}^H \mathbf{F} ).
\end{align}\vspace{-0.6cm}

\bibliographystyle{IEEEtran}
\bibliography{myref}

\begin{thebibliography}{10}
\providecommand{\url}[1]{#1}
\csname url@samestyle\endcsname
\providecommand{\newblock}{\relax}
\providecommand{\bibinfo}[2]{#2}
\providecommand{\BIBentrySTDinterwordspacing}{\spaceskip=0pt\relax}
\providecommand{\BIBentryALTinterwordstretchfactor}{4}
\providecommand{\BIBentryALTinterwordspacing}{\spaceskip=\fontdimen2\font plus
\BIBentryALTinterwordstretchfactor\fontdimen3\font minus
  \fontdimen4\font\relax}
\providecommand{\BIBforeignlanguage}[2]{{%
\expandafter\ifx\csname l@#1\endcsname\relax
\typeout{** WARNING: IEEEtran.bst: No hyphenation pattern has been}%
\typeout{** loaded for the language `#1'. Using the pattern for}%
\typeout{** the default language instead.}%
\else
\language=\csname l@#1\endcsname
\fi
#2}}
\providecommand{\BIBdecl}{\relax}
\BIBdecl

\bibitem{Li}
H.~Li, Y.~Liu, Y.~Chen, and Z.~Pan, ``Near-field integrated sensing,
  positioning, and communication,'' in \emph{Proc. {IEEE} Intl. Conf. Commun.
  (ICC)}, Jun. 2024, Submitted.

\bibitem{dang2020should}
S.~Dang, O.~Amin, B.~Shihada, and M.-S. Alouini, ``What should 6{G} be?''
  \emph{Nat. Electron.}, vol.~3, no.~1, pp. 20--29, 2020.

\bibitem{kraus2002antennas}
J.~D. Kraus and R.~J. Marhefka, ``Antennas for all applications,''
  \emph{Antennas for all applications}, 2002.

\bibitem{10220205}
Y.~Liu, Z.~Wang, J.~Xu, C.~Ouyang, X.~Mu, and R.~Schober, ``Near-field
  communications: A tutorial review,'' \emph{IEEE open j. Commun. Soc.},
  vol.~4, pp. 1999--2049, 2023.

\bibitem{1137900}
J.~Sherman, ``Properties of focused apertures in the fresnel region,''
  \emph{IEEE Trans. Antennas Propag.}, vol.~10, no.~4, pp. 399--408, 1962.

\bibitem{zhang20196g}
Z.~Zhang, Y.~Xiao, Z.~Ma, M.~Xiao, Z.~Ding, X.~Lei, G.~K. Karagiannidis, and
  P.~Fan, ``6{G} wireless networks: Vision, requirements, architecture, and key
  technologies,'' \emph{IEEE Veh. Technol. Mag.}, vol.~14, no.~3, pp. 28--41,
  2019.

\bibitem{rajatheva2020white}
N.~Rajatheva, I.~Atzeni, E.~Bjornson, A.~Bourdoux, S.~Buzzi, J.-B. Dore,
  S.~Erkucuk, M.~Fuentes, K.~Guan, Y.~Hu \emph{et~al.}, ``White paper on
  broadband connectivity in 6{G},'' \emph{arXiv preprint arXiv:2004.14247},
  2020.

\bibitem{5571900}
Y.~Shen and M.~Z. Win, ``Fundamental limits of wideband localization— part
  {I}: A general framework,'' \emph{{IEEE} Trans. Inf. Theory}, vol.~56,
  no.~10, pp. 4956--4980, 2010.

\bibitem{zhang2022beam}
H.~Zhang, N.~Shlezinger, F.~Guidi, D.~Dardari, M.~F. Imani, and Y.~C. Eldar,
  ``Beam focusing for near-field multiuser {MIMO} communications,''
  \emph{{IEEE} Trans. Wireless Commun.}, vol.~21, no.~9, pp. 7476--7490, 2022.

\bibitem{9620081}
H.~Lu and Y.~Zeng, ``Near-field modeling and performance analysis for
  multi-user extremely large-scale {MIMO} communication,'' \emph{{IEEE} Commun.
  Lett.}, vol.~26, no.~2, pp. 277--281, 2022.

\bibitem{10123941}
Z.~Wu and L.~Dai, ``Multiple access for near-field communications: {SDMA} or
  {LDMA}?'' \emph{{IEEE} J. Sel. Areas Commun.}, vol.~41, no.~6, pp.
  1918--1935, 2023.

\bibitem{xiao2022overview}
Z.~Xiao and Y.~Zeng, ``An overview on integrated localization and communication
  towards 6{G},'' \emph{Sci. China Inf. Sci.}, vol.~65, pp. 1--46, 2022.

\bibitem{wang2011new}
G.~Wang and K.~Yang, ``A new approach to sensor node localization using {RSS}
  measurements in wireless sensor networks,'' \emph{{IEEE} Trans. Wireless
  Commun.}, vol.~10, no.~5, pp. 1389--1395, 2011.

\bibitem{wang2015asymptotically}
Y.~Wang and K.~Ho, ``An asymptotically efficient estimator in closed-form for
  3-{D} {A}o{A} localization using a sensor network,'' \emph{{IEEE} Trans.
  Wireless Commun.}, vol.~14, no.~12, pp. 6524--6535, 2015.

\bibitem{qiao2023sensing}
L.~Qiao, A.~Liao, Z.~Li, H.~Wang, Z.~Gao, X.~Gao, Y.~Su, P.~Xiao, L.~You, and
  D.~W.~K. Ng, ``Sensing user's activity, channel, and location with near-field
  extra-large-scale {MIMO},'' \emph{arXiv preprint arXiv:2307.10837}, 2023.

\bibitem{wang2023cram}
H.~Wang, Z.~Xiao, and Y.~Zeng, ``Cram\'er-{R}ao bounds for near-field sensing
  with extremely large-scale {MIMO},'' \emph{arXiv preprint arXiv:2303.05736},
  2023.

\bibitem{hua2023near}
H.~Hua, J.~Xu, and Y.~C. Eldar, ``Near-field 3{D} localization via {MIMO}
  radar: {C}ram\'er-{R}ao bound analysis and estimator design,'' \emph{arXiv
  preprint arXiv:2308.16130}, 2023.

\bibitem{wang2023near}
Z.~Wang, X.~Mu, and Y.~Liu, ``Near-field integrated sensing and
  communications,'' \emph{{IEEE} Commun. Lett.}, 2023.

\bibitem{cong2023near}
J.~Cong, C.~You, J.~Li, L.~Chen, B.~Zheng, Y.~Liu, W.~Wu, Y.~Gong, S.~Jin, and
  R.~Zhang, ``Near-field integrated sensing and communication: Opportunities
  and challenges,'' \emph{arXiv preprint arXiv:2310.01342}, 2023.

\bibitem{10273424}
X.~Zhang, H.~Zhang, and Y.~C. Eldar, ``Near-field sparse channel representation
  and estimation in 6{G} wireless communications,'' \emph{{IEEE} Trans.
  Commun.}, pp. 1--1, 2023.

\bibitem{7942128}
K.~T. Selvan and R.~Janaswamy, ``Fraunhofer and {F}resnel distances: Unified
  derivation for aperture antennas,'' \emph{IEEE Antennas Propag. Mag.},
  vol.~59, no.~4, pp. 12--15, 2017.

\bibitem{6717211}
O.~E. Ayach, S.~Rajagopal, S.~Abu-Surra, Z.~Pi, and R.~W. Heath, ``Spatially
  sparse precoding in millimeter wave {MIMO} systems,'' \emph{{IEEE} Trans.
  Wireless Commun.}, vol.~13, no.~3, pp. 1499--1513, 2014.

\bibitem{9598863}
X.~Wei and L.~Dai, ``Channel estimation for extremely large-scale massive
  {MIMO}: Far-field, near-field, or hybrid-field?'' \emph{{IEEE} Commun.
  Lett.}, vol.~26, no.~1, pp. 177--181, 2022.

\bibitem{kay1993fundamentals}
S.~M. Kay, \emph{Fundamentals of statistical signal processing: estimation
  theory}, Prentice-Hall, Inc., 1993.

\bibitem{zhang2006schur}
F.~Zhang, \emph{The Schur complement and its applications}, Springer Science \& Business Media, 2006, vol.~4.

\bibitem{9120361}
Q.~Shi and M.~Hong, ``Penalty dual decomposition method for nonsmooth nonconvex
  optimization—part {I}: Algorithms and convergence analysis,'' \emph{{IEEE}
  Trans. Signal Process.}, vol.~68, pp. 4108--4122, 2020.

\bibitem{grant2014cvx}
M.~Grant and S.~Boyd, ``{CVX}: Matlab software for disciplined convex
  programming, version 2.1,'' [Online]. Available:\url{http://cvxr.com/cvx},
  2014.

\bibitem{9124713}
X.~Liu, T.~Huang, N.~Shlezinger, Y.~Liu, J.~Zhou, and Y.~C. Eldar, ``Joint
  transmit beamforming for multiuser {MIMO} communications and {MIMO} radar,''
  \emph{{IEEE} Trans. Signal Process.}, vol.~68, pp. 3929--3944, 2020.

\bibitem{5447068}
Z.-q. Luo, W.-k. Ma, A.~M.-c. So, Y.~Ye, and S.~Zhang, ``Semidefinite
  relaxation of quadratic optimization problems,'' \emph{IEEE Signal Process.
  Mag.}, vol.~27, no.~3, pp. 20--34, 2010.

\bibitem{8359308}
X.~Zhang, W.~Chen, W.~Zheng, Z.~Xia, and Y.~Wang, ``Localization of near-field
  sources: A reduced-dimension {MUSIC} algorithm,'' \emph{{IEEE} Commun.
  Lett.}, vol.~22, no.~7, pp. 1422--1425, 2018.

\bibitem{6834753}
M.~R. Akdeniz, Y.~Liu, M.~K. Samimi, S.~Sun, S.~Rangan, T.~S. Rappaport, and
  E.~Erkip, ``Millimeter wave channel modeling and cellular capacity
  evaluation,'' \emph{{IEEE} J. Sel. Areas Commun.}, vol.~32, no.~6, pp.
  1164--1179, 2014.

\end{thebibliography}

\end{document}